\RequirePackage{rotating}
\documentclass[12pt]{iopart}
\usepackage[T1]{fontenc}       % Use modern font encodings
\usepackage{multirow}
\usepackage{graphicx}
\usepackage{rotating}
\usepackage{xcolor}
\expandafter\let\csname equation*\endcsname\relax 
\expandafter\let\csname endequation*\endcsname\relax 
 \usepackage{amsmath} 

\definecolor{deepjunglegreen}{rgb}{0.0, 0.29, 0.29}
\definecolor{vbred}{rgb}{1.0, 0.0, 0.0}
\definecolor{mrpurple}{rgb}{0.5, 0.0, 0.5}

\begin{document}

%\title[Organic Films Level Alignment]{Predicting Carrier Level Alignment in Large Hybrid Inorganic-Organic Systems by Hybrid Density Functional Theory: Tetracene and Pentacene at H/Si(111)}
%\title[Organic Films Level Alignment]{Pentacene and Tetracene on H/Si(111): Predicted Carrier Level Alignment for Large Low-Strain Interfaces from Hybrid Density Functional Theory}

\title[Pc and Tc on H/Si(111)]{Pentacene and Tetracene Molecules and Films on H/Si(111): Level Alignment from Hybrid Density Functional Theory}

\author{Svenja M. Janke$^{1,2}$, Mariana Rossi$^2$\footnote{Present address: Max Planck Institute for Structure and Dynamics of Matter, Hamburg, Germany}, Sergey V. Levchenko$^{3,2}$, Sebastian Kokott$^2$, Matthias Scheffler$^2$, Volker Blum$^{1,4}$}

\address{$^1$ Duke University, Mechanical Engineering and Materials Science, Duke University, Durham, NC, USA}
\address{$^2$ Fritz Haber Institute of the Max Planck Society, Berlin, Germany}
\address{$^3$Skolkovo Institute of Science and Technology, Moscow, Russia}
\address{$^4$Duke University, Department of Chemistry, Duke University, Durham, NC, USA}
%\ead{svenja.janke@duke.edu}

\begin{abstract}
The electronic properties of hybrid organic-inorganic semiconductor interfaces depend strongly on the alignment of the electronic carrier levels in the organic/inorganic components. 
In the present work, we address this energy level alignment from first principles theory for two paradigmatic organic-inorganic semiconductor interfaces, the singlet fission materials tetracene and pentacene on H/Si(111), using all-electron hybrid density functional theory. For isolated tetracene on H/Si(111), a type I-like heterojunction (lowest-energy electron and hole states on Si) is found. For isolated pentacene, the molecular and semiconductor valence band edges are degenerate. For monolayer films, we show how to construct supercell geometries with up to 1,192 atoms, which minimize the strain between the inorganic surface and an organic monolayer film. Based on these models, we predict the formation of type II heterojunctions (electron states on Si, hole-like states on the organic species) for both acenes, indicating that charge separation at the interface between the organic and inorganic components is favored. The paper discusses the steps needed to find appropriate low-energy interface geometries for weakly bonded organic molecules and films on inorganic substrates from first principles, a necessary prerequisite for any computational level alignment prediction.
\end{abstract}
\maketitle

\section{Introduction}
\begin{figure}[h]
\centering
\includegraphics[width=0.99\textwidth]{{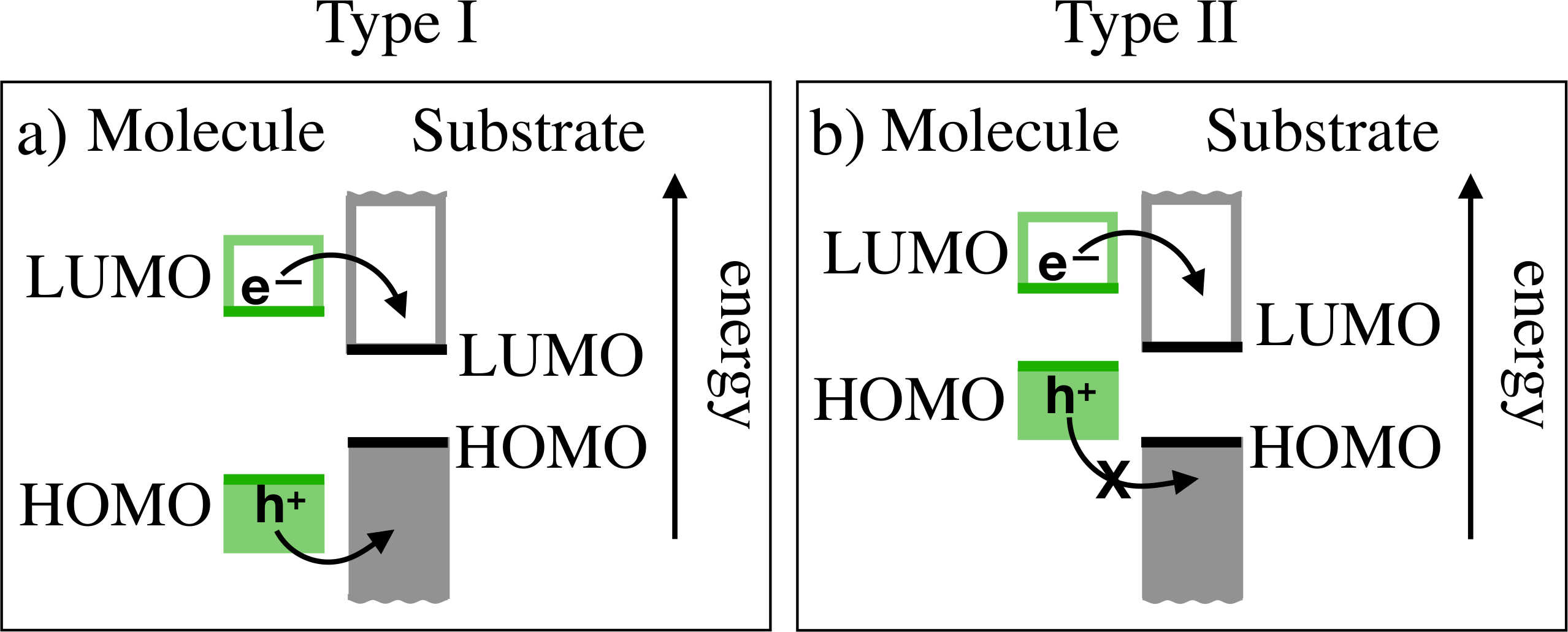}}
\caption{\label{Fig:heterojunction1}Two possible schematic energy level alignment diagrams for hybrid organic-inorganic systems. a) Type I heterojunction, where the organic film has a wider band gap than the inorganic substrate. b) Type II heterojunction, where the energy levels of the organic component are staggered with respect to the inorganic substrate.}
\end{figure}

The electronic level alignment between two different semiconductors in contact with each other is of significant technological and physical importance, determining phenomena such as internal charge separation, quantum confinement, or charge recombination in semiconductor heterostructures. A process of particular interest is the generation of excitons in one part of a heterostructure, which may dissociate into individual charged carriers that can be separated at an internal interface. In single-junction solar cells, charge carriers are collected at the band edges of semiconductors, i.e., the conduction band minimum (CBM) and the valence band maximum (VBM). As a result, the energy fraction of each absorbed photon that surpasses the energy of the band gap is usually lost as heat. This energy loss defines the Shockley-Queisser limit\cite{shockley1961} that limits the theoretical maximum efficiency of single-junction solar cells to below approximately 30~\%. A possible way to overcome the thermalization energy loss is singlet fission\cite{raoFriendNatRevMater2017, smith2010, congreve2013, allardiceRaoJACS2019, einzingerBaldoNature2019}. In singlet fission, a high-energy photon generates a singlet exciton, which can dissociate into two lower-energy triplet excitons prior to separation into individual carriers, resulting in four instead of two carriers and preserving a significant fraction of the energy that would otherwise be lost as heat.  Two prominent examples for fission materials are pentacene\cite{burgos1977, jundt1995} (Pc) and tetracene\cite{merrifieldGroffChemPhysLett1969, geacintov1969, swenbergStacyChemPhysLett1968, groff1970} (Tc), whose triplet exciton energies roughly match the band gap of silicon\cite{smith2010, ehrlerNatCom2012, jundt1995, wilson2013}. 

In principle, the Shockley-Queisser limit could be overcome by augmenting conventional solar cells with layers of singlet fission materials that enable charge carrier insertion from the triplet excitons into Si\cite{dexterJLum1979, smith2010, raoFriendNatRevMater2017, futscherEhrlerACSEnergyLett2018}. To transfer or split excitons at an organic/inorganic interface, such as Tc and Pc at H/Si(111), the  energy level alignment between the ``highest occupied molecular orbital'' and ``lowest unoccupied molecular orbital'' (HOMO and LUMO, respectively, here used synonymously with the VBM and CBM) of the components at the interface is a defining quantity\cite{tkatchenkoSchefflerMRSBull2010,tabachnykRaoNatMater2014, zhuKahnMRSBull2010}. Figure~\ref{Fig:heterojunction1} schematically illustrates two possible energy level alignments between an organic film and an inorganic substrate. Assuming sufficiently small exciton binding energies, a photogenerated triplet exciton in the organic film could either dissociate into two carriers that cross into the substrate at the heterojunction (type I, Figure~\ref{Fig:heterojunction1}~a) or split into a pair of carriers, of which only one enters the substrate (type II, Figure~\ref{Fig:heterojunction1}~b)\cite{raoFriendNatRevMater2017, dexterJLum1979}. A type I level alignment has been proposed for Tc on H/Si(111) although hole extraction from H/Si(111) to Tc was still observed in the same study\cite{macqueenLipsMaterHoriz2018}. Direct triplet insertion from Tc into passivated Si appears to be at least hindered~\cite{pilandCPL2014, macqueenLipsMaterHoriz2018, futscherEhrlerACSEnergyLett2018}. For Pc, based on electron affinity and ionization potential measurements, type II heterojunction behavior was suggested\cite{campbellCroneJApplPhys2009}.

\begin{figure}[ht]
\centering
\includegraphics[width=0.80\textwidth]{{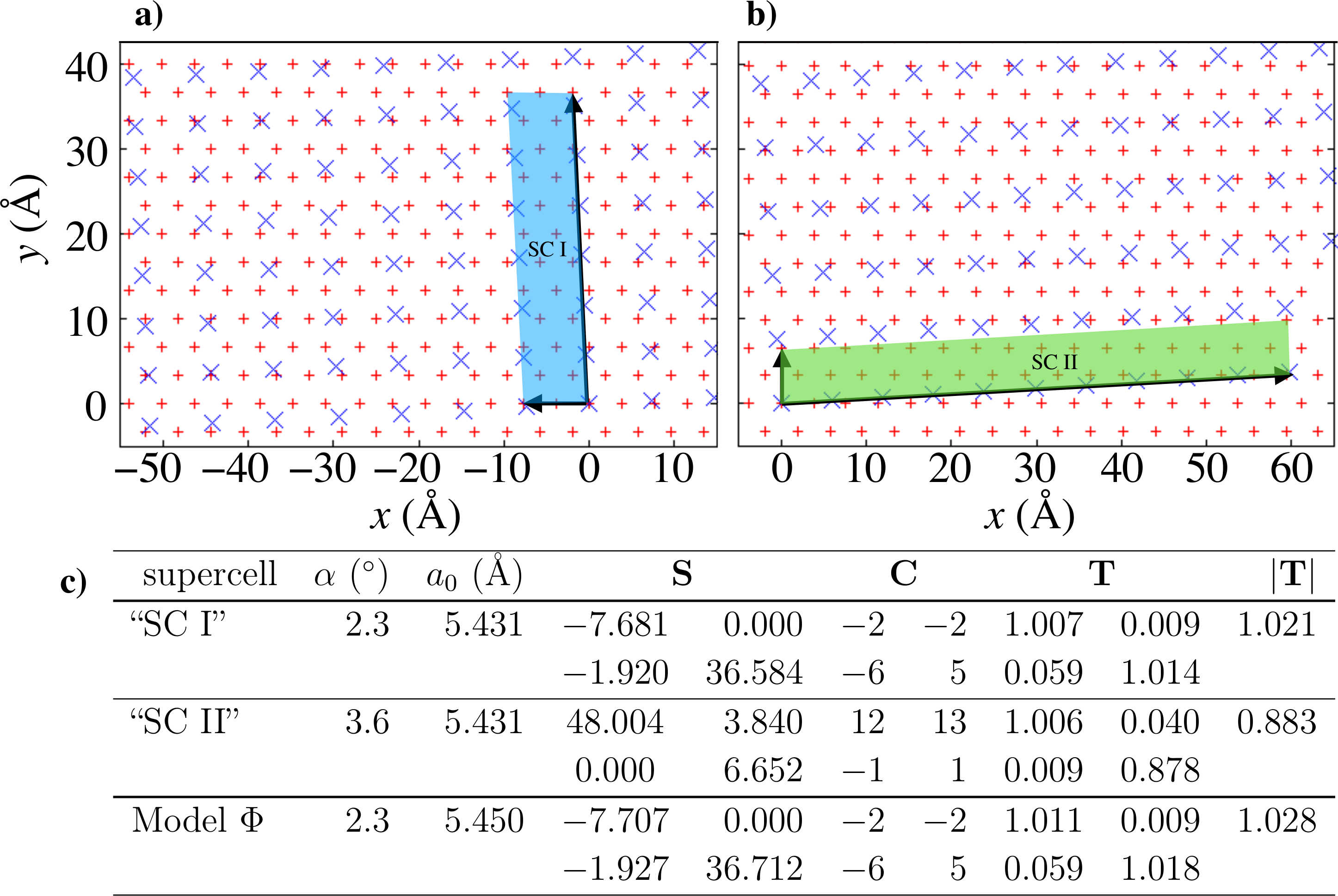}}
\caption{\label{Fig:Pc:Nishikata:epitaxialmatrix}Overlay of the substrate and film lattices based on the experimental Si lattice parameter (5.431~\AA\cite{kittel2005}, red crosses) and a) the Pc film unit cell alignment I (blue crosses) and b) the Pc film unit cell alignment II (blue crosses) proposed by Nishikata \textit{et al.}\cite{nishikataPRB2007}. Experimental lattice parameters for the unstrained film unit cells I and II are reproduced in Table~\ref{Tab:experimental:lattice parameters} further below. Hypothetical film supercells (SCs) that have been strained to match the substrate lattice are shown as blue (SC I) and green (SC II) areas, showing that this can be accomplished by a relatively small strain for SC I vs. a rather large strain for SC II. c) Numerical representation of the strained film SCs shown in subfigures a) and b) that lead to exact coincidence of the supercell lattice vectors with the underlying substrate lattice. $\alpha$ is the rotational angle of the Pc film on the substrate and $a_0$ is the assumed lattice parameter of the substrate. Eqs. \ref{Eq:S:epitaxy}, \ref{Eq:film:epitaxy} and \ref{Eq:strain} in Sec. ``Computational Approach'' define the in-plane lattice parameter matrix of the combined film-substrate supercell, $\mathbf{S}$, the supercell matrix $\mathbf{C}$ with respect to the substrate's (111) plane and the strain transformation $\mathbf{T}$ from the unstrained (incommensurate) to the strained (commensurate) film. $|\mathbf{T}|$ summarizes the area strain between the unstrained\cite{nishikataPRB2007} and the strained film supercell. $\mathbf{S}$ and $\mathbf{T}$ for ``SC I'' and ``SC II'' are based on the experimental substrate lattice parameter of Si and the experimental film lattice parameters of Ref.\cite{nishikataPRB2007}.  ``Model $\Phi$'' is the same model as SC I but its $\mathbf{S}$ is defined using the computational lattice parameter for bulk Si (DFT-PBE+TS level of theory, see below).}
\end{figure}

In this paper, we predict the electronic level alignment of Pc and Tc molecules and monolayer films on intrinsic, i.e., undoped H/Si(111) using first principles theory, specifically using hybrid density-functional theory. For the correct description of this alignment\cite{tkatchenkoSchefflerMRSBull2010, zhuKahnMRSBull2010, heimelKochAdvFunctMater2009,
wangRossiAdvElMater2019, akaikeJJAplPhys2018} and consequently the singlet fission properties\cite{gishJohnsonJPCC2019}, building a model unit cell that reflects the geometric structure at the interface is crucial. However, the resulting unit cells are large and computationally demanding. On the one hand, the interaction between Tc or Pc and H/Si(111) is weak, as evidenced\cite{hlawacekTeichert2013} by the observed lack of a wetting layer\cite{shimada2005} and the standing in-plane (herringbone) orientation of Pc\cite{nishikataPRB2007} and Tc\cite{shi2006, tersigni2006} molecules around room temperature in thin films and islands on surfaces like H/Si(111). On the other hand, the unit cell lattice parameters for islands, monolayers and thin films of both Pc on H/Si(111)\cite{nishikataPRB2007, ruiz2003, shimada2005, sadowski2008} and Pc and Tc at other weakly interacting surfaces\cite{fritzToneyJACS2004, yangBaoJACS2005, tersigni2006, tersigni2011,  matsubara2011, nabokDraxl2007,  ruizPRL2003,  schiefer2007, shi2006, verlaakDeleuzePRB2003, yoshidaSato2007, ruizIslamApplPhysLett2004,  sadowskiTromp2005} do not match the unit cell lattice parameter expected\cite{kittel2005} for H/Si(111) well. As a result\cite{hooksWardAdvMater2001}, large low-strain commensurate approximate unit cells are necessary, leading to slab models of the interface with hundreds of atoms or more. Indeed, Pc molecules on H/Si(111) grow in at least two distinct orientations\cite{nishikataPRB2007}, which require large unit cells to approximate them, as shown in Figure~\ref{Fig:Pc:Nishikata:epitaxialmatrix} and further discussed in Section \ref{sec:results}. 

%\begin{sidewaystable}
%%\begin{table}
%\centering
%\begin{tabular}{lrrcrrrrrrrr} 
% \hline\
%supercell & $\alpha$ ($^\circ$)& $a_0$ (\AA) &  \multicolumn{2}{c}{$\mathbf{S}$}& \multicolumn{2}{c}{$\mathbf{C}$} & \multicolumn{2}{c}{$\mathbf{T}$}& $|\mathbf{T}|$\\
%\hline
%\hline	
%``SC I'' &	2.3&5.431	&$-7.681$ &0.000&$-2$&	$-2$&1.007&	0.009&	1.021\\
%%0.991&	0.040&	0.980\\
%&&&	$-1.920$&$36.584$&		$-6$&	5&	0.059&	1.014\\
%%0.012&	0.989	\\
%\hline
%``SC II''&	3.6&5.431&$48.004$&3.840&		12&	13&1.006&	0.040&	0.883	\\
%%1.000&	$-0.016$&	1.133\\
%&&&	0.000&6.652&	$-1$&	1&0.009&	0.878\\
%%	$-0.071$&	1.134	\\
%
%\hline
% \hline
%Model $\Phi$&	2.3&5.450&$-7.707$ &0.000	&$-2$&	$-2$&1.011&	0.009&	1.028 \\
%%	0.988&	0.040&	0.973\\
%&&&$-1.927$&$ 36.712$&		$-6$&	5&0.059	&1.018 \\
%%	0.012&	0.985\\
%\hline
%\end{tabular}
%%\end{table}
%\end{sidewaystable}

The need for large commensurate supercells aggravates the already challenging determination of level alignments from first principles. It is well established that the electronic delocalization errors\cite{cohenYangChemRev2012, sanchezYangPRL2008} associated with the relatively affordable level of density-functional theory (DFT) in the generalized gradient approximation (GGA) lead to fundamental band gaps that are too small, sometimes incorrect ordering of electronic levels and hence potentially qualitatively wrong level alignments\cite{maromJPhysCondMatter2017, maromKronikJChemPhys2008, koerzdoerferKronikPRB2009, 
koerzdoerferKronikPRB2010, MaromRinkePRB2012}. Methods that offer significant improvement over DFT-GGA, like the quasi-particle $GW$ approximation\cite{tamblynNeatonPhysRevB2011, maromJPhysCondMatter2017, draxlHannewaldAccChemRes2014} or hybrid functionals in DFT\cite{garzaScuseriaJPhysChemLett2016, hofmannSchefflerJCP2013, perdewYangPNAS2017}, are associated with high computational cost that either make them very demanding or, in the case of $GW$, essentially prohibit\cite{maromJPhysCondMatter2017, garzaScuseriaJPhysChemLett2016, hofmannSchefflerJCP2013} their application to systems of the size required here. 

A recent, deeper discussion of the double challenge of potentially large structure model sizes on the one hand and of achieving a sufficently high level of theory to capture all relevant effects that affect energy level alignments on the other hand can be found, e.g., in Ref.~\cite{Nabok2019}. In that work, a nearly strain-free interface (poly(para-phenylene)at the rock-salt ZnO(100) surface) was investigated in detail, allowing for small model sizes and thus an analysis of the electronic level alignment up to the $G_0W_0$ level of theory. In contrast, the present work focuses on analyzing the construction of nearly strain-free computational structure models in a system where a small-cell, low-strain approximate structure is not available. Electronic effects are analyzed at the already rather challenging level of hybrid DFT, which offers the appropriate mathematical form to capture at least highest occupied and lowest unoccupied levels in principle\cite{perdewYangPNAS2017} and which can be extended to a form that accounts for simple consequences of screening as well.\cite{kronikkuemmelAdvMater2018}

In the following, we address the task of creating suitable low-strain supercells for interfaces between Pc or Tc and H/Si(111). Two types of organic-inorganic system geometries were investigated for this work: (i) molecules adsorbed at H/Si(111) in isolation from one another, referred to as the ``dilute limit'', and (ii) molecules forming a closed monolayer-like film on H/Si(111), referred to as the ``monolayer limit'' below. By combining these models with hybrid density functional theory calculations, we arrive at a fully computational approach for predicting the structure and level alignment between these acene films and the H/Si(111) substrate.

\section{\label{sec:comput}Computational Approach}
\subsection*{Computational Details}
All DFT calculations were carried out using the all-electron electronic structure code FHI-aims\cite{blum2009, havuSchefflerJCompPhys2009, renSchefflernNewJPhys2012, knuthSchefflerCPComm2015} with large scale calculations facilitated by the ELSI\cite{elsiYuBlum2018} infrastructure, the ELPA eigenvalue solver\cite{elpaMarekLederer2014}, and a linear-scaling implementation of hybrid functionals in periodic DFT\cite{Ihrig_2015,levchenkoSchefflerCompPhysComm2015}. For structure prediction, we used DFT-GGA in the Perdew-Burke-Ernzerhof (PBE) exchange-correlation functional\cite{PBE11996} together with the Tkatchenko-Scheffler (TS) pairwise dispersion scheme\cite{tkatchenko2009}. In combination with PBE, the TS scheme has been shown to reproduce lattice vectors and volumes of organic crystals closely\cite{hojaTkatchenkoWIRE2017, alSaidiJordanJCTC2012, bedoyamartinezZojerJCTC2018} and predicts lattice parameters and internal geometries within 2~\% of experimental results for acene bulk materials\cite{SchatschneiderTkatchenkoPRB2013}. In the present work, we find similarly good agreement for crystal polymorphs of Pc and Tc, as well as for the lattice constant for bulk Si, as shown in the supplementary material, Table~S1, Table~S2 and Table~S3 (these tables also include a comparison to a more recent many-body dispersion (MBD) scheme\cite{ambrosetti2014}). 

Hybrid density functionals include a fraction of non-local exact exchange that partially corrects the delocalization error. Compared to $GW$, hybrid DFT provides a more affordable balance between accuracy and computational cost, while retaining the appropriate mathematical form\cite{perdewYangPNAS2017} to yield acceptable fundamental gaps for typical semiconductors\cite{garzaScuseriaJPhysChemLett2016, hofmannSchefflerJCP2013}. We investigated the electronic level alignments using the Heyd-Scuseria-Ernzerhof (HSE06) functional with $\alpha = 25$~\% Hartree-Fock exchange and a screening parameter of $\omega=0.11$~(Bohr radii)$^{-1}$\cite{HSEJChemPhys2003, HSEJChemPhys2006, krukauScuseria2006} by single-point calculations using DFT-PBE+TS predicted geometries. Full and projected electronic densities of states (PDOS) were computed using a Gaussian broadening function with a width of 0.1~eV.  The energetic positions of the frontier levels of the organic and inorganic subsystems were extracted from a Mulliken analysis\cite{mullikenJCP1955}, on which the PDOS computations are also based. %counting all contributions above 0.0025. 
Visualizations of atomic configurations were obtained using the Jmol\cite{jmol} and VMD\cite{vmd} computer programs. 
 
Adsorption energies of acene molecules on the substrate are calculated as
\begin{equation}
\label{Eq:Epermol}
\Delta E = \frac{E_\mathrm{Sub + Mol}^{n\times m} - n\times m \cdot E_\mathrm{Sub}^{1\times 1} - N_\mathrm{Mol}\cdot E_\mathrm{Mol}}{N_\mathrm{Mol}} .
\end{equation}
$E_\mathrm{Sub + Mol}^{n\times m}$ is the total energy of the combined molecules-on-substrate model, $N_\mathrm{Mol}$ is the number of molecules, and $n\times m$ is the number of atoms per layer in the H/Si(111) substrate. $E_\mathrm{mol}$ and $E_\mathrm{Sub}^{1\times 1}$  are the total energy of the isolated molecule and of a ($1\times 1$) unit cell of the H/Si(111) substrate in vacuum, respectively.  If multiple molecules are present in a cell, we calculate the average of descriptors of the structure (the absorption distance $d_z$, angle with the surface normal $\theta$ and the herringbone angle $\omega$ between the molecules).

\subsection*{H/Si(111)}
The lateral unit cell parameter of all final system geometries involving H/Si(111) slabs was set to $a_{111} = 3.854$~{\AA}, derived from the predicted lattice parameter of bulk Si, $a_0 = $5.450~{\AA}, using DFT-PBE+TS and ``tight'' settings (see Table~S3 in the supplementary material for the variation of $a_0$ with different density functionals). All slab geometries involving H/Si(111) were hydrogenated on both sides. Only one side was decorated with molecules for models of the organic films. A dipole correction\cite{neugebauerSchefflerPRB1992} was used to minimize any residual interaction between slab surfaces across the vacuum.

To determine acceptable but affordable slab thicknesses of H/Si(111), we investigated the convergence of the substrate electronic frontier levels as a function of slab thickness without adsorbates. These simulations were carried out using vacuum layers of 200~{\AA} between the slabs, (12$\times$12$\times$1) k-point grids, and FHI-aims' ``tight'' computational defaults, i.e., benchmark-quality settings\cite{LejaeghereScience2016, huhnBlumPRMater2017}. The H positions and outermost two Si double layer atomic positions were fully relaxed at the DFT-PBE+TS level of theory. As shown in Figure~S1 in the SI, the energetic positions of the VBM and CBM converge slowly with the number of layers in the slab, similar to other findings in the literature\cite{liGalliPRB2010, sagisakaBowlerJPhyCondensMat2017, delleySteigmeierApplPhysLett1995, scherpelzGalliPRMater2017} and attributed to quantum well behavior due to confinement of the electronic eigenstates in the thin slab\cite{scherpelzGalliPRMater2017, delleySteigmeierApplPhysLett1995, fischettiJiseok2011}. For the larger DFT-HSE06 supercell calculations including organic films, we use ten- and six-double-layer slabs. As shown in Figure~S1, for six double-layers, the CBM calculated by DFT-HSE06 is still approximately 0.2~eV higher than for thick slabs. Similarly, the calculated band gap of the six-double-layer slab is 1.465~eV, approximately 0.3~eV higher than the calculated bulk band gap. The slow convergence of the gap with slab thickness and the computational cost for DFT-HSE06 calculations of structures above 1,000~atoms (the largest film models considered in this work using six-double-layer slabs) make it impossible to systematically consider much thicker slab models. We return to this point below, concluding that the expected remaining CBM shift for a thicker slab would not be large enough to alter the qualitative film-substrate level alignments resulting from our calculations. 

\subsection*{The Dilute Limit}
For the dilute limit of acene adsorption, we placed single Tc molecules in $(4\times 4)$ supercells of the H/Si(111) substrate, whereas Pc molecules were placed in $(5\times 5)$ supercells. As shown in Figure~S2 in the supplementary material, the size of the supercells is sufficient to isolate the molecules from their periodic images.

\begin{figure}[t]
\centering
\includegraphics[width=0.99\textwidth]{{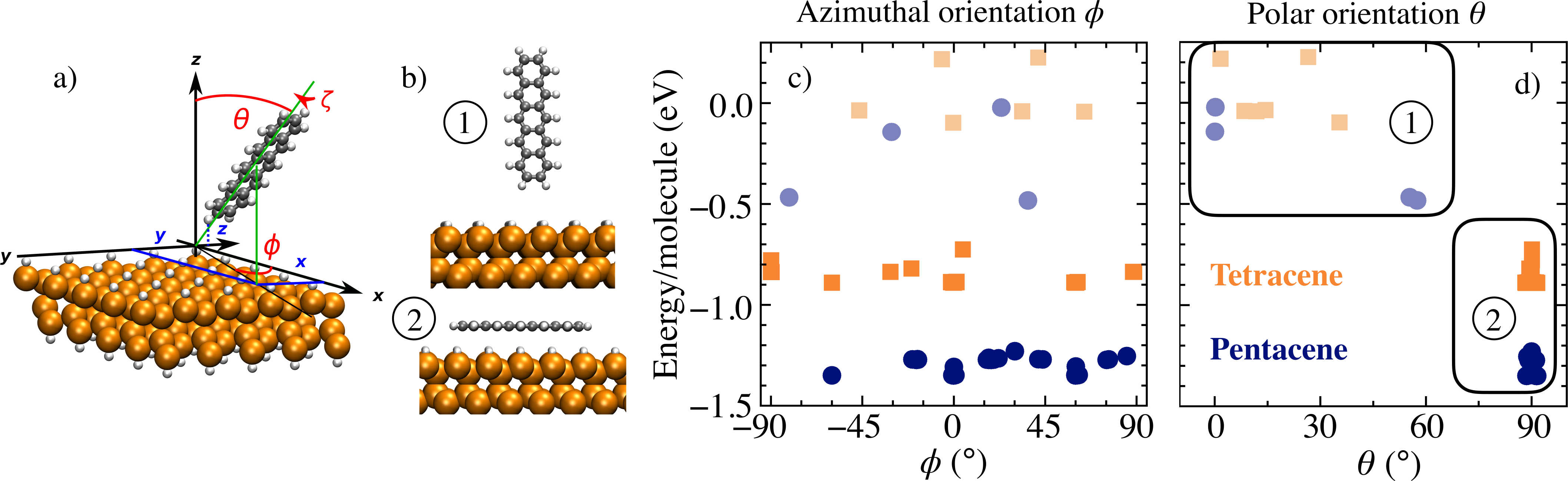}}
\caption{\label{Fig:PcTc:dilute:molor}a) Degrees of freedom of a rigid Tc molecule on H/Si(111); the center of mass is at $x$, $y$ and $z$. The molecule's orientation is given by the azimuth angle $\phi$ of the molecule's long axis with the $x$-axis ($[1\bar{1}0]$-direction), the polar angle $\theta$ of the molecule's long axis with the $z$-axis and the rotation of the molecule around its long axis by the angle $\zeta$. b) Example of a molecule in a standing (1) and in a lying (2) geometry. After geometry relaxation with DFT-PBE+TS, the calculated energy per molecule (Eq.~\ref{Eq:Epermol}) for different geometries of Pc (blue  circles) and Tc (orange squares) is given as a function of c) the azimuth angle and d) the polar angle. Lighter shades indicate a standing orientation (1), darker shades a lying orientation (2).} 
\end{figure}

As illustrated in Figure~\ref{Fig:PcTc:dilute:molor}a, adsorption geometries can be characterized by the molecular orientation with respect to the surface and by the molecule's lateral placement. Different local minima of the potential energy surface (lying vs. standing geometries, see Figure~\ref{Fig:PcTc:dilute:molor}b) were determined from a total of 145 initial geometry starting points each for Pc and Tc. 45 starting geometries were selected according to a grid of $\theta = 0^\circ$, $45^\circ$ and $90^\circ$ with $\phi = 30^\circ$, $60^\circ$ and $90^\circ$.  For $\theta = 90^\circ$, rotation around the molecule's long axis $\zeta = 0^\circ$, $45^\circ$ and $90^\circ$ were sampled. The molecules' centers of mass were placed at arbitrary $x$- and $y$-positions at 2~{\AA} above the plane of the H atoms. The remaining 100 starting geometries were selected by orienting the molecules randomly in the ranges $\theta = 0^\circ$ -- $90^\circ$, $\phi = -30^\circ$ -- $90^\circ$ and placing the molecules' centers of mass at randomly chosen $x$- and $y$-positions within the supercell, at vertical positions between $z = 1.6$ -- $4.0$~{\AA}. 

Structure optimization was initially carried out using FHI-aims' ``light'' settings and the DFT-PBE+many-body dispersion (MBD)\cite{ambrosetti2014} approach on two double layered H-terminated Si slabs. This initial set of pre-optimized geometries was next refined using the PBE+TS scheme for consistency with other simulations in this work. We chose a subset of 50 configurations for refined relaxations, consisting of the thirty lowest-energy configurations plus twenty picked from the rest of the initial pool. The latter also included configurations where the acene molecule was found to be standing on the substrate. For these configurations, the slab thicknesses were increased to ten double-layers. During post-relaxation using FHI-aims' ``intermediate'' settings and DFT-PBE+TS, the slabs were separated by 65~{\AA} of vacuum and the top four double layers were allowed to move. After post-relaxation, all residual forces were below 0.005~eV/\AA. DFT-HSE06 follow-up calculations of electronic total and projected densities of states were carried out using ``intermediate'' settings. The k-point meshes employed for relaxation and electronic structure investigations of the different structures are detailed in Table~S4 in the supplementary material.

\subsection*{The Monolayer Limit}
We pursued two different approaches to obtain suitable low-strain models for Pc/Tc monolayer films on the H/Si(111) substrate:
\begin{itemize}
\item Based on experimentally determined\cite{nishikataPRB2007} film lattice parameters and the periodicity at room temperature (see Figure \ref{Fig:Pc:Nishikata:epitaxialmatrix}), we built a computational model for Pc monolayer films on H/Si(111), called ``Model $\Phi$''.
\item As an alternative approach, we used a protocol to obtain combined film-on-substrate supercells, independent of whether experimental lattice parameters and periodicity are known. In this protocol, geometries for the ``monolayer limit'' were obtained by fitting freestanding monolayer Tc and Pc films onto the H/Si(111) substrate. Details of this construction methodology are reported in the results section below.
\end{itemize}

\begin{figure}[t]
\centering
\includegraphics[width=0.99\textwidth]{{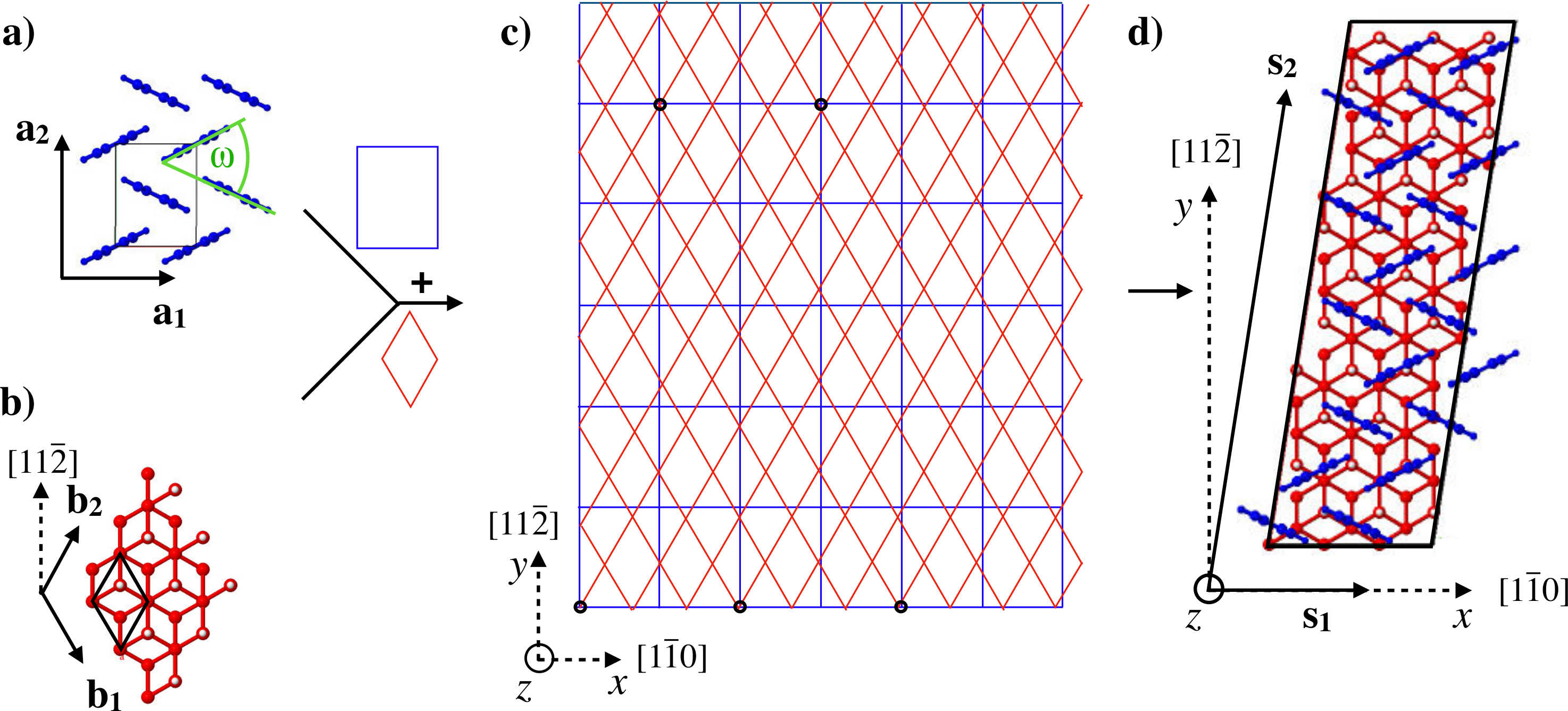}}
\caption{\label{Fig:DenseFilm}a) The in-(001)-plane unit cell of the acene film (blue) with the lattice parameters $\mathbf{a_1}$ and  $\mathbf{a_2}$ and the herringbone angle $\omega$ between the molecules. b) The H/Si(111) surface unit cell with lattice parameters $\mathbf{b_1}$ and $\mathbf{b_2}$ as well as the [11$\bar{2}$] direction. Si atoms are depicted in red, H atoms in  white. c) Extended unit cell representations of film and substrate, superimposed to find approximate points of coincidence (black). d) Example coincidence supercell constructed from points of approximate coincidence (model C). The $x$, $y$ and $z$-axes of the coordinate system as well as two surface directions of the H/Si(111) substrate and the lattice parameters of the combined system $\mathbf{s_1}$ and $\mathbf{s_2}$ are shown.}
\end{figure}

For the second construction strategy, initial two-dimensional lattice parameters for free-standing monolayer film models (no substrate) were found by fully relaxing their lateral unit cells using DFT-PBE+TS, $(10\times 10\times 1)$ k-point meshes, and FHI-aims' ``intermediate'' settings. Adjacent monolayers were separated by at least 75~{\AA} of vacuum. The free-standing monolayer geometries consist of ``standing'' molecules arranged side by side in a herringbone pattern. For Pc, this herringbone structure is modeled after the experimentally known geometries\cite{nishikataPRB2007} of Pc films on H/Si(111) at room temperature. For Tc, they are modeled from a separately constructed unit cell with two Tc molecules arranged in a herringbone pattern (see Figure~\ref{Fig:DenseFilm} as well as Figure~S3 and Table~S5 in the SI). The final relaxed DFT-PBE+TS lattice parameters of Pc and Tc films are almost identical but slightly shorter than the experimentally reported lattice parameters for Pc monolayers on weakly interacting surfaces (see discussion in the results section).

Commensurate models of combined film/substrate supercells are described as follows. For a weak film/substrate interaction, the energetic disadvantage of straining the film (needed to form an exact coincidence lattice) will outweigh the advantage of adsorbing molecules at energetically preferred adsorption sites on the substrate\cite{hooksWardAdvMater2001}. Thus, we consider film-substrate geometries based on adsorbed films that are only minimally strained compared to a free-standing model. In a mathematical representation, the lattice vectors of the combined film/substrate supercell used in simulations are multiples of the substrate lattice vectors $\mathbf{b}_1$ and $\mathbf{b}_2$ (the lattice vectors defining the grid of red crosses in Figures \ref{Fig:Pc:Nishikata:epitaxialmatrix}a and b). The supercell lattice vectors $\mathbf{s}_1$ with their individual components ($s_{11}$, $s_{12}$) and $\mathbf{s}_2$ with ($s_{21}$, $s_{22}$) can be expressed in matrix notation as follows\cite{hooksWardAdvMater2001}:
\begin{equation}
\label{Eq:S:epitaxy}
\left(
\begin{matrix}
\mathbf{s}_1\\
\mathbf{s}_2
\end{matrix}
\right)  =
\mathbf{C} \cdot
\left(
\begin{matrix}
\mathbf{b}_1\\
\mathbf{b}_2
\end{matrix}
\right)
=
\left(
\begin{matrix}
c_{11} & c_{12}\\
c_{21} & c_{22}
\end{matrix}
\right) \cdot
\left(
\begin{matrix}
\mathbf{b}_1\\
\mathbf{b}_2
\end{matrix}
\right)
\end{equation}
The coefficients $c_{11}$, $c_{12}$, $c_{21}$ and $c_{22}$ are integers.  Similarly, for primitive film lattice vectors $\mathbf{a_1}$ and $\mathbf{a_2}$ (the lattice vectors defining the grid of blue crosses in Figures \ref{Fig:Pc:Nishikata:epitaxialmatrix}a and b), we can define superlattice vectors $\mathbf{f}_1$ and $\mathbf{f}_2$ of the film, using different sets of integer coefficients $\tilde{c}_{11}$, $\tilde{c}_{12}$, $\tilde{c}_{21}$ and $\tilde{c}_{22}$:
\begin{equation}
\label{Eq:film:epitaxy}
\left(
\begin{matrix}
\mathbf{f}_1\\
\mathbf{f}_2
\end{matrix}
\right)  =
\mathbf{\tilde{C}} \cdot
\left(
\begin{matrix}
\mathbf{a}_1\\
\mathbf{a}_2
\end{matrix}
\right)
=
\left(
\begin{matrix}
\tilde{c}_{11} & \tilde{c}_{12}\\
\tilde{c}_{21} & \tilde{c}_{22}
\end{matrix}
\right) \cdot
\left(
\begin{matrix}
\mathbf{a}_1\\
\mathbf{a}_2
\end{matrix}
\right)
\end{equation}
A supercell of the low-strained film will have unit vectors $\mathbf{f}_1$ and $\mathbf{f}_2$ that are close to two substrate superlattice vectors $\mathbf{s}_1$ and $\mathbf{s}_2$. Accordingly, a strain transformation $\mathbf{T}$ of $\mathbf{f}_1$ and $\mathbf{f}_2$ can be introduced so that the resulting strained film supercell matches the substrate supercell exactly. Defining $\mathbf{s}_1$, $\mathbf{s}_2$ and $\mathbf{f}_1$, $\mathbf{f}_2$ as rows of two matrices $\mathbf{S}$ and $\mathbf{F}$, respectively, we can write: 
\begin{equation}
  \label{Eq:strain}
  \mathbf{S} = \mathbf{T}\mathbf{F}.
\end{equation}
Here, the film supercell vectors $\mathbf{F}$ can correspond to rotated or unrotated versions of the overall film with respect to the substrate. In either case, for low-strain approximant supercells, the matrix $\mathbf{T}$ should be as close as possible to the identity matrix. The determinant $|\mathbf{T}|$ is a measure of the area strain on the film. A value larger than unity corresponds to a stretched film, and a value lower than unity corresponds to compressive strain.

For atomic position optimization within large supercell models combining Pc or Tc monolayers with the H/Si(111) substrate, we employed the k-point meshes as shown in Table~S4 in the SI. Of the H/Si(111) slabs containing six double layers of Si, the top four Si double layers, the surface H~atoms and the molecular adsorbates were allowed to relax using DFT-PBE+TS and FHI-aims' ``intermediate'' settings until the residual forces on all optimized atoms were smaller than 0.005~eV/\AA. The resulting geometries were then used to calculate the electronic structure (total and projected densities of states) of Tc and Pc monolayers on H/Si(111) using DFT-HSE06, FHI-aims' ``intermediate'' settings and the k-space meshes detailed in Table~S4 in the SI.  

\section{Results \label{sec:results}}
\subsection{\label{Subsec:Res:DiluteLimit}Dilute Limit}
In Figure~\ref{Fig:PcTc:dilute:molor}c and Figure~\ref{Fig:PcTc:dilute:molor}d, we show the adsorption energies $\Delta E$ (Eq.~\ref{Eq:Epermol}) of individual Tc and Pc molecules on H/Si(111) as a function of the azimuthal angle $\phi$ of the molecule's long axis with the $x$-axis (Figure~\ref{Fig:PcTc:dilute:molor}c) and $\theta$ of the molecule's long axis with the surface normal (Figure~\ref{Fig:PcTc:dilute:molor}c). Each data point corresponds to a specific, fully relaxed geometry obtained from a different starting geometry. Neither Pc nor Tc show a distinct preference for a particular azimuthal orientation (Figure~\ref{Fig:PcTc:dilute:molor}c). In contrast, both acene molecules prefer a lying orientation (i.e. $\theta \approx 90^\circ$, $\zeta = 90^\circ$ in Figure~\ref{Fig:PcTc:dilute:molor}d and conformation (2) in Figure~\ref{Fig:PcTc:dilute:molor}b) over a standing orientation (i.e. $\theta < 60^\circ$ in Figure~\ref{Fig:PcTc:dilute:molor}d and conformation (1) in Figure~\ref{Fig:PcTc:dilute:molor}b). This agrees with previous experimental\cite{shimada2005} and computational\cite{tsetseris2005, ample2008, diLabio2009} observations of Pc on weakly interacting surfaces and with the general observation that a lying, rod-like, aromatic molecule should have a stronger interaction with a substrate than a standing one\cite{hlawacekTeichert2013}. %The mean adsorption distance of lying Tc and Pc is $d_z = 4.7$~{\AA} and $4.6$~{\AA}, respectively, with a minimum energy per molecule of $\Delta E = -0.890$~eV for Tc and $\Delta E = -1.350$~eV for Pc. 
The minimum energy per molecule is $\Delta E = -0.890$~eV for lying Tc and $\Delta E = -1.350$~eV for lying Pc. Among the cases we investigated, only four of the Pc and six of the Tc cases assume a standing orientation.  The minimum energies per molecule found for standing Tc ($-0.097$~eV) and for standing Pc ($-0.483$~eV) are much less favorable than for the lying case. Two standing Tc molecules display $\Delta E > 0.0$~eV, indicating that, if the molecules are initially placed too far from the substrate ($d_z>11.3$~{\AA}), the interaction between molecules and substrate may not be sufficiently large to relax into local minima based on the minimum force criteria chosen here.

\begin{figure}
\centering
\includegraphics[width=0.8\textwidth]{{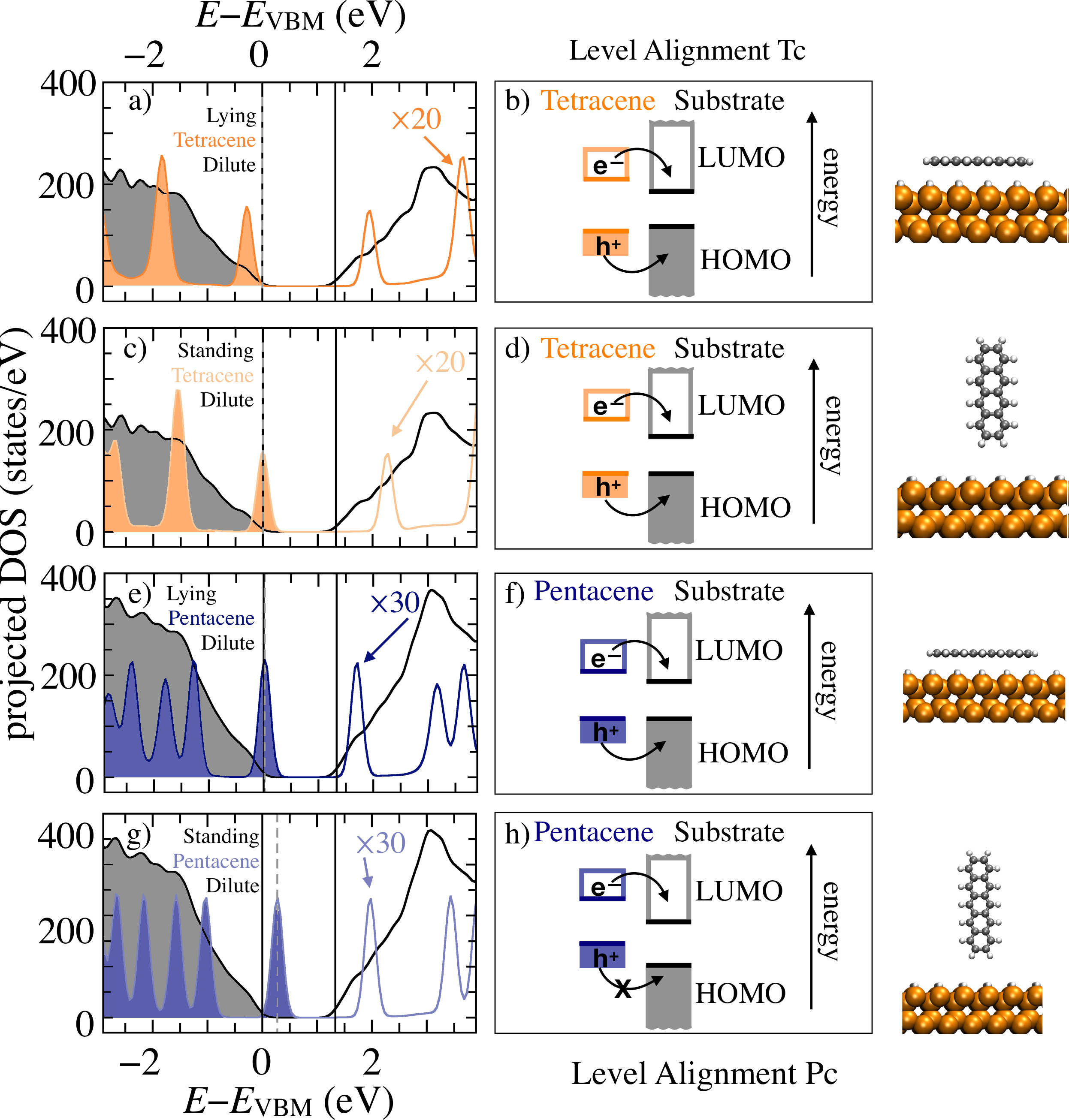}}
\caption{\label{Fig:PcTc:dilute:film:PDOS}DFT-HSE06 projected densities of state (PDOS) for single Tc and Pc molecules adsorbed on H/Si(111) aligned to the substrate's VBM, decomposed into PDOS for the H/Si(111) substrate atoms (black curves) and the adsorbed Tc (orange curves) and Pc (blue curves) molecules. Filled areas indicated occupied levels, empty areas unoccupied levels. Vertical solid lines indicate the positions of inorganic HOMO and LUMO, the dashed grey line marks the position of the electron chemical potential.  Due to the artificial Gaussian broadening (0.1~eV) applied to compute the PDOS, the peaks of the overall HOMO appear broadened beyond the actual position of the chemical potential (the structures shown are in fact insulating, not metallic).
a) PDOS for the lying Tc case. b) Schematic energy level diagram of the HOMOs and LUMOs of the lying Tc case, c) PDOS for the standing Tc case. d) Schematic energy level diagram of the standing Tc case, e) PDOS for the lying Pc case. f) Schematic energy level diagram of the lying Pc case, g) PDOS for the standing Pc case, h) Schematic energy level diagram of the standing Pc case. The orientation of the organic molecule on the substrate is illustrated schematically on the right. Because the molecule contains much fewer atoms than the underlying substrate, the DOS projected onto the acene films were scaled by a factor of 20 (Tc) and 30 (Pc) to make them comparable to the DOS projected onto the substrate atoms. Note that the supercells chosen for Pc (5$\times$5) contain significantly more Si atoms per molecule than those chosen for Tc (4$\times$4).}  
\end{figure}

In Figure~\ref{Fig:PcTc:dilute:film:PDOS}, we visualize the DFT-HSE06 predicted electronic densities of states of the energetically most favorable ``dilute'' Tc and Pc adsorption geometries (for both lying and standing cases), projected onto the isolated molecules and onto the substrate, as well as schematic depictions of the frontier level alignment for each case. For lying Tc (Figure~\ref{Fig:PcTc:dilute:film:PDOS}a and b), the organic HOMO and LUMO fall into the substrate's valence and conduction band (type I alignment). This means that in principle, charge transfer from the molecule to the substrate should be possible for both hole- and electron-like carriers. For standing Tc (Figure~\ref{Fig:PcTc:dilute:film:PDOS}c and d), the molecular HOMO and substrate VBM are practically degenerate. Lying Pc (Figure~\ref{Fig:PcTc:dilute:film:PDOS}e and f) also shows a degenerate molecular HOMO and substrate VBM, whereas for standing Pc (Figure~\ref{Fig:PcTc:dilute:film:PDOS}g and h) the molecular HOMO would be in the substrate gap (type II alignment). A transfer of hole carriers from the molecule to the substrate would be hindered. 

From Figure~\ref{Fig:PcTc:dilute:film:PDOS} we observe that, from the lying to the standing orientation, the molecular frontier levels shift almost uniformly upwards by $\sim 0.25$~eV compared to the substrate bands. The shift is due to different electrostatic interactions between the standing or lying molecule and the surface\cite{heimelKochAdvFunctMater2009}. In the lying case, the molecule's $\pi$-system interacts more strongly with the underlying substrate. 
The HOMO and LUMO are hence shifted downward in energy compared to the standing case. Note that varying the hybrid functional\cite{skoneGalliPRB2016} or considering dynamical screening effects\cite{neatonHybertsenPRL2006, refaelyabramsonKronikPRB2013,Nabok2019} might alter the energy levels further. Nevertheless, the results indicate that hole transfer from  isolated molecules to the substrate should be more difficult for a standing molecule than for a lying one.

\subsection{Dense Monolayer Limit}
\label{Sec:Dense}
\begin{sidewaystable}
%\begin{table}
\caption{\label{Tab:experimental:lattice parameters}Experimental lattice parameters $\mathbf{a_1}$ and $\mathbf{a_2}$ for acene films on weakly interacting surfaces from selected literature sources, compared to free-standing monolayer film models in vacuum (i.e., without including a substrate) after full relaxation by DFT-PBE+TS in this work. $\gamma$ is the angle between the lateral lattice vectors. RT is room temperature, GIXRD is grazing-incidence x-ray diffraction, XRD is x-ray diffraction, LEED is low-energy electron diffraction, RHEED is reflection high energy electron diffraction, and STM is scanning tunneling microscopy.}
\begin{tabular}{lllllllll} 
 \hline
Molecule & Coverage &		$\mathbf{a}_1$ (\AA)&	$\mathbf{a}_2$ (\AA)&$\gamma (^\circ)$&	Substrate& Method& Temperature &Ref.\\
\hline\hline
%Tc&1~ML&5.92&7.33&90&free-standing&&&this work\\
Tc&1~ML&5.85&7.39&90&free-standing&&&this work\\
\hline
Tc&3~ML&$5.5\pm0.6$&$7.3\pm0.6$&&H/Si(001)&STM&RT&\cite{tersigni2006}\\
Tc&1.5-12~ML&$5.93\pm0.02$&$7.70\pm0.02$&90.0$\pm0.2$&H/Si(001)&GIXRD&RT&\cite{tersigni2011}\\
\hline
\hline
Pc &1~ML&5.85&7.36&90&free-standing&&&this work\\
\hline
Pc&1~ML& 6.32$\pm0.06$&7.73$\pm0.07$&84$\pm1$&H/Si(111)&RHEED &65$^\circ$C& \cite{shimada2005}\\
Pc&$<1$~ML&6.02$\pm0.01$&7.62$\pm0.02$&90.0$\pm0.4$&H/Si(111)&LEED& RT&unit cell I\cite{nishikataPRB2007}\\
Pc&$<1$~ML&5.98$\pm0.01$&7.56$\pm0.01$&90.3$\pm0.1$&H/Si(111)&LEED& RT&unit cell II\cite{nishikataPRB2007}\\
\hline
Pc &1~ML &6.1$\pm0.2$&7.89&86$\pm0.5$&Bi(001)&LEED&RT&\cite{sadowskiTromp2005}\\
Pc&1~ML&5.916&7.588&89.95&SiO$_2$&GIXRD&RT&\cite{fritzToneyJACS2004}\\
Pc &$<1$~ML&5.90 & 7.62& 90$\pm0.2$&SiO$_2$&GIXRD&0, 22, 45$^\circ$C&\cite{ruizIslamApplPhysLett2004}\\

%Pc& 180~nm&5.92 & 7.54 & 89.9 & SiO$_2$ &GIXRD & 60$^\circ$C&~\onlinecite{nabokDraxl2007}\\
%Pc&50~nm &5.93 & 7.56 & 89.8& SiO$_2$&RSM&&~\onlinecite{yoshidaSato2007}\\
Pc&0.6--100~nm &5.94& 7.54& 89.5& SiO$_2$&GIXRD& 20$^\circ$C, 50$^\circ$C&\cite{kakudateSaitoApplPhysLett2007}\\
%Pc &$48$~nm & 5.958$\pm0.005$& 7.596$\pm0.008$&89.80$\pm0.10$&SiO$_2$, OTS, Topas &GIXRD &RT &\onlinecite{schiefer2007}\\
%Pc& 19~nm&5.91& 7.58&90$\pm0.2$&SiO$_2$&GIXRD&22$^\circ$&\onlinecite{ruizIslamApplPhysLett2004}\\
Pc&2~ML&5.90$\pm0.01$&7.51$\pm0.01$&89.92$\pm0.01$&SAMs&GIXRD& RT&\cite{yangBaoJACS2005}\\
Pc&$<30$~nm& 5.9 & 7.5 & 90 & KCl(001)&XRD& RT&\cite{kiyomuraIsodaJapJApplPhys2006}\\
%Pc & & 6.1$\pm0.1$&7.6$\pm0.1$&89.5$\pm$0.9&NaCl(100)&ED & RT &\onlinecite{wuSpenceJApplCrhst2004}\\
%Pc & & 6.2$\pm0.1$&7.7$\pm0.1$&84.8$\pm0.9$&NaCl(100)&ED & 80$^\circ$C &\onlinecite{wuSpenceJApplCrhst2004}\\
\hline
\end{tabular}
%\newline $^\mathrm{a}$ Unit cell I, $^\mathrm{b}$  unit cell II
\end{sidewaystable}
%\end{table}

For the monolayer films, we first consider theoretical models based on the experimentally observed coincidence pattern for Pc on H/Si(111)\cite{nishikataPRB2007}. In a second step, we test a protocol to obtain approximate combined film-on-substrate supercells, independent of whether experimental lattice parameters and periodicity are known.

\subsubsection{Interface Model Based on Experimental Lattice Coincidence (``Model $\Phi$'')}
 Nishikata \textit{et al.}\cite{nishikataPRB2007} identified point-on-line coincidence of superlattices of Pc-dendrites with the H/Si(111) substrate lattice at room temperature from low-energy electron microscopy (LEEM) and low-energy electron diffraction (LEED). Pc was observed to grow in two orientations on the H/Si(111) substrate with slightly different unit cells, labelled ``I'' and ``II'' in Figure~\ref{Fig:Pc:Nishikata:epitaxialmatrix} and  Table~\ref{Tab:experimental:lattice parameters}. Nishikata \textit{et al.} characterized supercells based on the film lattice parameters and their orientation. For supercells ``I'', a periodicity of ($36.0\pm 0.8$)~\AA~(six unit vectors of the Pc film) was identified in the $\mathbf{a}_1$-direction.  For supercells ``II'', the identified periodicity is ($47.84\pm0.01$)~{\AA} (eight unit vectors of the Pc film) in the $\mathbf{a}_1$-direction. Both supercells show a periodicity of one unit vector of the Pc film in the $\mathbf{a}_2$-direction\cite{nishikataPRB2007}.
Using Eqs. \ref{Eq:S:epitaxy}, \ref{Eq:film:epitaxy} and \ref{Eq:strain}, we can evaluate the experimentally reported supercells. The results are tabulated in Figure \ref{Fig:Pc:Nishikata:epitaxialmatrix}c. For supercell ``I'', we find $|\mathbf{T}| = 1.021$. This amounts to a stretch of the film on the substrate with a resulting area strain on the film of 2~\%. For supercell ``II'', $|\mathbf{T}| = 0.883$. A 12~\% area compression is large for a weakly interacting system. We conclude that supercell ``II'' cannot easily be modeled as commensurate structure with the underlying H/Si(111) substrate, unless a much larger commensurate supercell in the $\mathbf{a}_2$ direction is considered. Additionally, Nishikata \textit{et al.}\cite{nishikataPRB2007} reported that supercell ``I'' is more common than supercell ``II''. We therefore focus our comparison on supercell ``I'' with lattice vectors based on the experimentally suggested periodicity (see Figure~\ref{Fig:Pc:Nishikata:epitaxialmatrix}c).

We first investigate the experimentally observed coincidence pattern for Pc on H/Si(111) for supercell ``I''. Based on the film lattice parameters determined by Nishikata \textit{et al.}\cite{nishikataPRB2007} and the DFT-PBE+TS silicon lattice parameter of $a_0=5.450$~{\AA} (to ensure a strain-free H/Si(111) substrate in the computations) the area strain of the computational model is characterized by $|\mathbf{T}|$=1.028. 
This combination is referred to as ``Model $\Phi$'' in Figure~\ref{Fig:Pc:Nishikata:epitaxialmatrix}c. After geometry relaxation with DFT-PBE+TS, the calculated adsorption energy per Pc molecule is $\Delta E = -2.222$~eV. As expected, the film is energetically more favorable than the adsorption of an isolated, lying Pc molecule on the substrate due to more favorable molecule-molecule interactions within the film (see also Table~S5 in the supplementary material). The molecules form a herringbone pattern with a computed ``edge-to-face'' angle of $\omega=49^\circ$ between the planes of neighboring molecules, and with a tilt of $\theta=22^\circ$ to the surface normal (see Figure~\ref{Fig:PcTcHSi111:coincidencepattern}a). %The average adsorption distance defined as the average of a molecule's center-of-mass and the first layer of Si-atoms in H/Si(111) is 9.47~{\AA}. 

Regarding experimental values for $\omega$ and $\theta$ in monolayer films, we are not aware of an ``apples-to-apples'' comparison for a Pc monolayer on H/Si(111). However, a value of $\omega=52.7^\circ$ was reported in a grazing-incidence X-ray diffraction (GIXRD) study of Pc monolayer films on amorphous silicon oxide\cite{mannsfeldBaoAdvMater2009}. Several different phases of Tc and Pc are known to be stable at room temperature, with reported herringbone angles both computationally and experimentally in a broadly similar range\cite{fritzToneyJACS2004,yoshidaSato2007,nabokDraxl2007,schiefer2007,mannsfeldBaoAdvMater2009,meyenburgHeimbrodtPCCP2016, siegristGordon2001}. 
The same GIXRD study\cite{mannsfeldBaoAdvMater2009} as well as a GIXRD study of two layers of Pc on self-assembling membranes\cite{yangBaoJACS2005} place $\theta$ at $\sim0$--$4^\circ$. This is similar to the tilt angle proposed for the so-called thin-film phase of Pc\cite{yoshidaSato2007,schiefer2007, ambroschdraxlMeisenbichlerNewJPhys2009} whose lattice parameters are similar to those of the monolayer\cite{yoshidaSato2007}.
In contrast, our value of $\theta=22^\circ$ is closely in line with tilt angles in the range of 20$^\circ$ to 28$^\circ$ observed in bulk-like Pc polymorphs, both experimentally\cite{yoshidaSato2007, holmesVollhardt1999, campbell1961} and computationally\cite{nabokDraxl2007, dellavalleGirlandoChemPhysChem2009}.
Refs.\cite{mannsfeldBaoAdvMater2009} and \cite{haasSiegristPRB2007} discuss the discrepancy of the tilt angle in the monolayer in terms of reduced film lattice parameters compared to the bulk phase and, therefore, a higher in-plane molecular density with corresponding upright molecules. Another mechanism that could lead to more upright molecules in experiment than those in our fully relaxed geometries is thermal motion. However, while past experimental studies of Pc polymorphs show some change of the tilt angle with temperature, the magnitude of the effect is in the range of a few degrees\cite{haasSiegristPRB2007} and does not support a thermally driven transition from  $\theta=22^\circ$ to fully upright molecules. Instead, we show below that computational film models using the DFT-PBE+TS density functional are denser and that nearly upright molecular geometries in the films, in line with GIXRD, would result for these denser computational film models as well.

\begin{figure}
\includegraphics[width=0.99\textwidth]{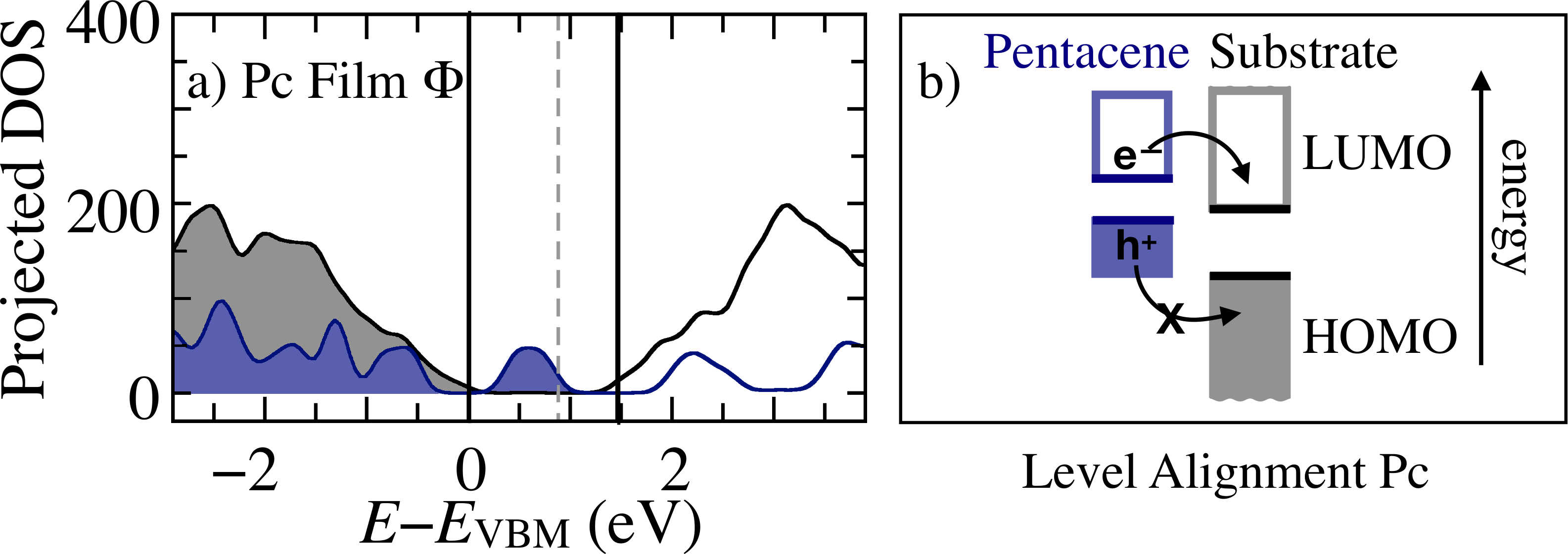}
\caption{\label{Fig:Pc:Phi:PDOS}a) DFT-HSE06 projected densities of states for the combined film-on-substrate model $\Phi$ based on the supercell ``I'' experimental lattice parameters and periodicity\cite{nishikataPRB2007}. Separate contributions are shown for the H/Si(111) substrate atoms (black) and the monolayer Pc molecules (blue). The solid black lines mark the positions of the substrate slab's HOMO (chosen as 0 eV) and LUMO and the dashed line that of the chemical potential. b) Schematic energy level diagram of the HOMO and LUMO of the H/Si(111) substrate (black) as well as the monolayer Pc film (blue) for the combined film-on-substrate model $\Phi$. In both panels, filled areas indicate occupied levels,  empty areas unoccupied  levels.}
\end{figure}

\subsubsection*{\label{Subsec:Phi}Electronic Properties of Model $\Phi$}
Figure~\ref{Fig:Pc:Phi:PDOS}a shows the DFT-HSE06 densities of states projected onto the substrate and Pc monolayer for the combined film-on-substrate model $\Phi$. A schematic energy level diagram is shown in Figure~\ref{Fig:Pc:Phi:PDOS}b. 
The LUMO of the adsorbed Pc monolayer is found within the substrate's conduction band. In contrast, the HOMO of the adsorbed Pc film is within the substrate's band gap. This type II level alignment is similar to Pc/Si alignment suggested based on separately  measured ionization potentials and electron affinities\cite{campbellCroneJApplPhys2009}. Based on this alignment, hole transfer from the film to the substrate would be hindered (holes would transfer from substrate to film instead), while electrons could still transfer into the substrate. 

Regarding the absolute values of the predicted gaps, the bulk band gap of Si ($\sim$1.17~eV at $T$=0~K\cite{kittel2005}) is well reproduced with the present flavor of DFT-HSE06 (calculated bulk band gap for ``tight'' settings, $\alpha = 25$~\%, $\omega=0.11$~(Bohr radii)$^{-1}$: 1.165~eV). As discussed above and shown in Figure~S2 in the supporting material, the band gap of a bare H/Si(111) slab of six Si double layers is still about 0.3~eV greater than the bulk band gap. The comparison to the vacuum level in Figure~S2 shows that the remaining change of the gap with increasing slab thickness is mainly due to the CBM, i.e., the qualitative band offsets between film and substrate reported based on six Si double layer slabs will likely remain unchanged since the substrate CBM would move down for a thicker slab. Importantly, in Figure ~\ref{Fig:Pc:Phi:PDOS}a the difference between substrate CBM (upper black solid line) and molecular HOMO (dashed grey line) is still significantly larger than 0.3~eV. Thus, the remaining overestimation of the substrate band gap due to the finite thickness of the slab is not expected to change the observed type-II level alignment concluded here for model $\Phi$.

For the pentacene film, the molecular orbitals are broadened compared to the dilute limit (Figure~\ref{Fig:PcTc:dilute:film:PDOS}). They are shifted up in energy compared to the isolated standing orientation, and the film's HOMO-LUMO gap is reduced to 1.05~eV. Qualitatively, both the broadening and the upwards shift are consistent with the increased confinement of the $\pi$-systems of the molecules in the more closely packed arrangement of the monolayer film. In comparison, the $G_0W_0$ band gap (using DFT in the local-density approximation as the starting point for $G_0W_0$) of a Pc solid is found to be 2.1~eV in Ref.~\cite{Refaely2015}, i.e., significantly larger than the DFT-HSE06 film gap predicted in the present work. While part of this difference may be attributable to structural differences in the film, the bulk of the discrepancy most likely stems from the different dielectric properties and environment of the Pc film compared to the Si bulk. For the Pc film, a different and higher $\alpha$ parameter in the HSE06 functional than for Si would be appropriate\cite{skoneGalliPRB2016}.

The difference between the predicted and the likely actual band gap of the Pc film is important when comparing qualitatively to the experimentally known excitonic properties of Pc. Experimentally, the emission energy of the  first triplet state in Pc was measured to be 0.81--0.90~eV\cite{geacintovStrom1971, burgos1977, vilarSchottChemPhysLett1983, deCheveigneDefourneauPRB1977}, a little less than half of the energy of the first singlet state\cite{wilson2013} of 1.83~eV\cite{sebastianBaesslerChemPhys1981}. The $G_0W_0$ gap for Pc\cite{Refaely2015} is correctly expected to be higher than the experimental singlet energy, whereas the DFT-HSE06 predicted gap in our film model is too low. However, given the computed considerable shift of the Pc HOMO into the substrate gap and given the expected downward shift of the actual substrate CBM compared to the slab model used in Figure~\ref{Fig:Pc:Phi:PDOS}, it is entirely conceivable that the experimentally expected triplet energy is still sufficient to allow for charge separation with holes remaining on the Pc film and electrons injected into the substrate.

\subsubsection{\label{Subsec:ComputationProcedure}Computational Procedure for Finding Lattice Coincidence Patterns}
We now turn to the determination of geometry and electronic level alignment between a film and a substrate by a purely computational approach, i.e., without relying on experimental input regarding the coincidence pattern. For Pc on H/Si(111), we can compare predictions to the results obtained above for ``Model $\Phi$''. For Tc on H/Si(111), the experimental coincidence pattern is unknown and the results presented below are thus our best available predictions for the level alignment in this system.

As described in Section~\ref{sec:comput}, we first relax free-standing monolayer films (i.e., \textit{in vacuo}) of acene molecules arranged in a herringbone pattern to obtain the lattice parameters that such films would assume without any interaction with the substrate. Table~\ref{Tab:experimental:lattice parameters} includes the predicted lateral lattice parameters and unit cell angles for the Tc and Pc model films in comparison to experimental unit cell parameters of Tc and Pc submonolayer films\cite{nishikataPRB2007, ruizIslamApplPhysLett2004}, monolayers\cite{shimada2005, sadowskiTromp2005, fritzToneyJACS2004} or thin films\cite{tersigni2006, tersigni2011, yangBaoJACS2005, kiyomuraIsodaJapJApplPhys2006, kakudateSaitoApplPhysLett2007} observed on different weakly interacting substrates. The experimentally reported unit cell angles $\gamma$ for Pc\cite{nishikataPRB2007} and Tc\cite{tersigni2006, tersigni2011} at room temperature on hydrogenated silicon are in good agreement with our simulated free-standing films. For the experimentally reported lattice parameters, there is noticeable scatter for Tc and Pc films on different substrates. Our theoretically determined Pc and Tc lattice parameters are within $<0.5$~\% of one another, i.e., essentially identical. Compared to the Pc monolayer films on H/Si(111)\cite{nishikataPRB2007} and to Tc thin films on H/Si(100)\cite{tersigni2011, tersigni2006}, the experimentally observed $\mathbf{a_2}$ parameters are larger by about 4~\%. The fully DFT-PBE+TS relaxed films are thus likely slightly denser than actual experimental Tc and Pc monolayer films. 

Figure~\ref{Fig:DenseFilm} illustrates the process of determining approximate coincidence lattices between the substrate and film superlattices for the case where the film $\mathbf{a_1}$ lattice vector is aligned with the $[1\bar{1}0]$ surface direction. The unit cells are shown for the acene film in Figure~\ref{Fig:DenseFilm}a and for the substrate unit cell in Figure~\ref{Fig:DenseFilm}b. Figure~\ref{Fig:DenseFilm}c shows the coincidence pattern between both lattices. Approximate points of close coincidence between lattice vectors of the acene and the H/Si(111) lattices are identified by black circles. A particular resulting combined cell, model C described below, is shown in Figure~\ref{Fig:DenseFilm}d.

\begin{figure}[t]
\centering
\begin{minipage}{1.0\textwidth}
\centering
\includegraphics[width=.99\textwidth]{{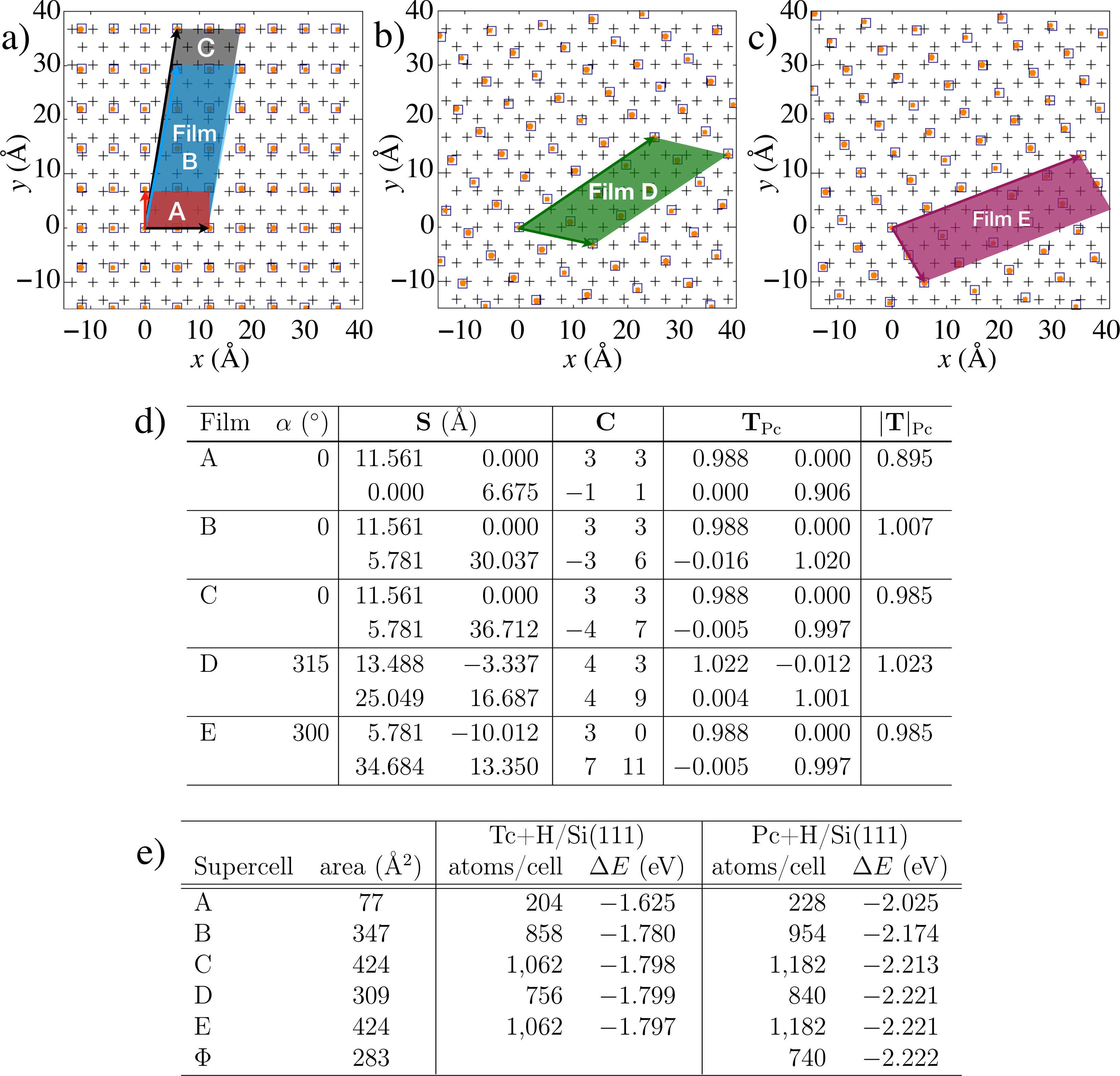}}
\end{minipage}
\caption{\label{Fig:PcTcHSi111:coincidencepattern}The coincidence pattern between the H/Si(111) substrate (black crosses), the tetracene film (orange circles) and the pentacene film (blue squares). Panels a--c) show different rotations $\alpha$ of the acene film with respect to the $[1\bar{1}0]$-direction of H/Si(111). a) $\alpha = 0^\circ$ for cells A (red), C (blue) and D (black) b) $\alpha = 315^\circ$ for unit cell B and c) $\alpha = 300^\circ$ for unit cell  E. Although cell C (a, black) and E (c, purple) have the same area, they are not symmetry equivalent: H/Si(111) exhibits three-fold symmetry due to the ABC stacking of the Si double layers. d) The matrix of the in-(001)-plane lattice vectors of the combined film-substrate supercell $\mathbf{S}$ is a multiple ($\mathbf{C}$) of the substrate lattice vectors $\mathbf{b}_1$ and $\mathbf{b}_2$. The strain transformation of the supercell parameters of the film $\mathbf{F}$ when the film is combined with the substrate is expressed in terms of the matrix $\mathbf{T}$. |$\mathbf{T}$| describes the agreement between the area of the substrate and monolayer film supercell. e) The combined (film plus substrate) unit cell areas, the number of atoms in the computational structure models, and the film adsorption energy per molecule calculated with DFT-PBE+TS for a Pc or Tc monolayer on H/Si(111).}  
\end{figure} 

We followed two different routes to construct combined film-on-substrate supercells. In the first route, we identified approximate coincidence points between substrate and film superlattices visually, by overlaying substrate and supercell lattice points. In the second route, described in more detail below, we obtained the points by a script written for this purpose. The script compares the agreement between possible multiples of unit cells for films and substrate. Because of the very similar lattice parameters predicted for Pc and Tc free-standing films, the same set of resulting commensurate supercells is used for Tc and for Pc in our simulations. In the following, we will discuss the lattice coincidence at the example of Pc. Similar conclusions apply to Tc.  
Figure~\ref{Fig:PcTcHSi111:coincidencepattern} shows unrotated and rotated commensurate approximant supercells between the substrate and the films, labelled as Models A-E, that were selected for further study in this work. Models A-C in Figure~\ref{Fig:PcTcHSi111:coincidencepattern}a illustrate different models from the first (visual) strategy, whereas models D and E in Figure~\ref{Fig:PcTcHSi111:coincidencepattern}b and c) were derived from the script-based strategy. The table in Figure~\ref{Fig:PcTcHSi111:coincidencepattern}d) summarizes their 2\,D lattice vectors and area strains as defined in Eq.~\ref{Eq:S:epitaxy} and Eq.~\ref{Eq:strain}. 
Finally, Figure~\ref{Fig:PcTcHSi111:coincidencepattern} lists their unit cell areas, number of atoms included in the full (film plus substrate) structure model, and DFT-PBE+TS calculated film adsorption energies per molecule for both Pc and Tc.

Model A, shown as the red area in Figure~\ref{Fig:PcTcHSi111:coincidencepattern}a, is the smallest possible commensurate unit cell that corresponds to reasonably close coincidence points of the lattices. While the coincidence in the $x$ direction is close, in the $y$ direction the molecular supercell is more extended than the corresponding underlying substrate supercell. In Figure~\ref{Fig:PcTcHSi111:coincidencepattern}d, both $\mathbf{T}$ and $|\mathbf{T}|$ reveal quantitatively that the film-substrate mismatch in the small-cell commensurate approximant model A corresponds to a compression of the film of $\sim10$~\% in $y$ direction. This is a significant compression that, as we show below, leads to a noticeable change of the electronic structure of the combined film-substrate model compared to larger coincidence cells.
Lower strain is achieved for the larger commensurate unit cells B and C, shown as the  blue and grey areas in Figure~\ref{Fig:PcTcHSi111:coincidencepattern}a. The values of $\mathbf{T}$ and $|\mathbf{T}|$ associated with them are much closer to unity. However, these models necessitate much larger overall numbers of atoms than A, and are  thus a challenge for electronic structure predictions beyond the level of hybrid functional DFT.

To obtain models D and E (Figure~\ref{Fig:PcTcHSi111:coincidencepattern}b and c), a script was used that explores the search space of points of coincidence systematically in two steps. 
\begin{itemize}
\item The first step tests the difference in area of possible substrate and film supercells $\mathbf{S}$ and $\mathbf{F}$ by evaluating $|\mathbf{T}|$. $\mathbf{S}$ was sampled from $\mathbf{C} = \bigl( \begin{smallmatrix} -10 & -10\\ -10 & -10 \end{smallmatrix}\bigr)$ to $\mathbf{C} = \bigl( \begin{smallmatrix} 10 & 10\\ 10 & 10 \end{smallmatrix}\bigr)$, covering 190,612 supercells and a maximum unit cell area of 1286~{\AA}$^2$. 
% 10x10,  only positive 6,961 supercells
% 9x9: 1042~\AA.
%11x11:9946 supercells and a maximum supercell area of 1556~\AA.
%12x12: 13754 supercells and a maximum unit cell area of 1852~\AA$^2$ (old script, w/o negatives). 
Because the lattice vectors $\mathbf{a}_1$ and $\mathbf{a}_2$ are almost twice as long as $\mathbf{b}_1$ and $\mathbf{b}_2$, we sampled corresponding film supercells $\mathbf{F}$ from $\mathbf{\tilde{C}} = \bigl( \begin{smallmatrix} -5 & -5\\ -5 & -5 \end{smallmatrix}\bigr)$ to $\mathbf{\tilde{C}} = \bigl( \begin{smallmatrix} 5 & 5\\ 5 & 5 \end{smallmatrix}\bigr)$, covering 
%563 
13,808 supercells and a maximum unit cell area of 1077~\AA$^2$. 
\item The second step rotates the film's $\mathbf{a_1}$ axis compared to the substrate's $\mathbf{b_1}$ axis by an angle $\alpha$.
If, in the first step, $|1-|\mathbf{T}|| < 1.1$, i.e., if the areas of $\mathbf{S}$ and $\mathbf{F}$ agree within 10~\%, $\mathbf{F}$ is rotated by $\alpha$ in steps of $1^\circ$ with respect to $\mathbf{S}$, resulting in a new transformation matrix $\mathbf{T}^{(\alpha)}$. After the second step, a total of %133,753 
159,890,112 possible supercells were compared for Pc. 
\end{itemize}

\begin{figure}[h]
   \centering
	\includegraphics[width=0.55\textwidth]{{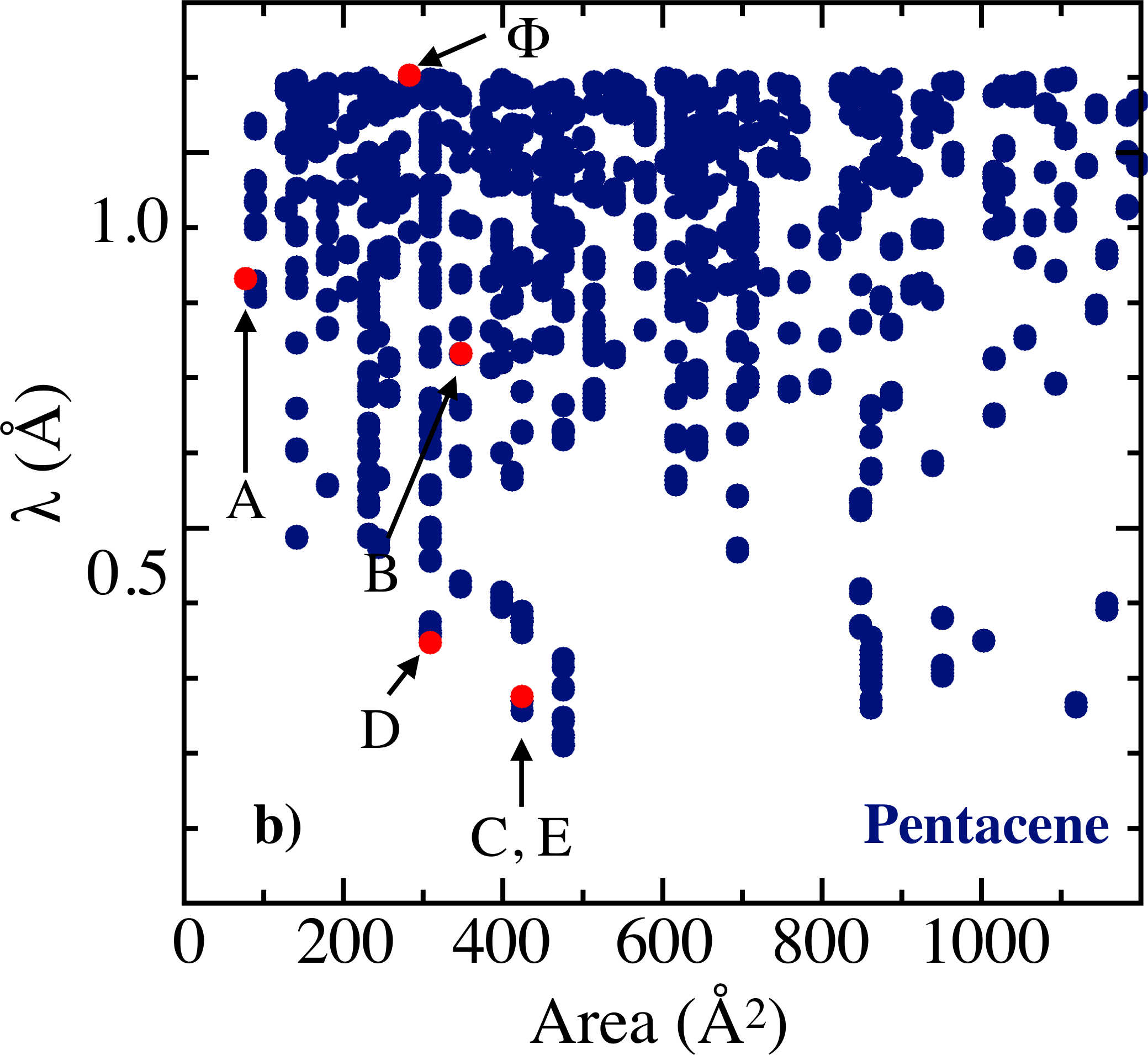}}
	\caption{\label{Fig:dilutelimit:lAcvsHSi111}
    Least-squares difference $\lambda$ between the lattice vectors of the H/Si(111) substrate and Pc  (blue circles) as a function of the H/Si(111) supercell area for different commensurate coincidence lattice models determined by a script as described in the text. Supercell models A-E, as well as the supercell model $\Phi$ determined according to the experimental study of Nishikata \textit{et al.}\cite{nishikataPRB2007} for Pc, are marked in red and indicated by arrows.}
\end{figure}

While $|\mathbf{T}|$ provides a good measure of the agreement of the area of the film and substrate supercell, this does not mean that the supercell lattice parameters of film and substrate agree well or match in shape. To find the best match between the supercell lattice parameters for a rotation $\alpha$, the agreement between the actual 2D lattice parameters of the film and substrate supercell for a rotation $\alpha$ can be quantified by the combination of substrate and film supercell lattice parameters that show the smallest disagreement:
\begin{equation}
\label{Eq:l1l2}
\lambda =  \sqrt{\sum_n^2 \min_{i = 1,2}\left(\sum_k^2 \left(f^{(\alpha)}_{n,k}-s_{i,k}\right)^2\right)}
\end{equation}
Figure~\ref{Fig:dilutelimit:lAcvsHSi111} categorizes different film-substrate models for Pc in terms of their coincidence supercell area ($x$-axis) as a measure of computational cost, and in terms of their lattice parameter mismatch $\lambda$ ($y$-axis) as a measure of degree of lattice coincidence for $\lambda<1.1$~{\AA}. The script-optimized models D and E display the lowest or close to lowest $\lambda$ values for their respective unit cell areas in Figure~\ref{Fig:dilutelimit:lAcvsHSi111}. The visually adjusted model C shows a low $\lambda$ value, whereas the differently stretched model B does not agree so well. Likewise the small-cell approximant A shows a rather high $\lambda$, in addition to its considerable area strain $|\mathbf{T}|$ in Figure~\ref{Fig:PcTcHSi111:coincidencepattern}d. As a final point of comparison, the experimental supercell I for Pc\cite{nishikataPRB2007}, Model $\Phi$ in its commensurate form is also shown in Figure~\ref{Fig:dilutelimit:lAcvsHSi111}b. Due to its size and the noticeable difference in lattice parameters between the freestanding Pc film model and the supercell I deduced from experiment, the $\lambda$ value of this model is considerably higher than for the specifically constructed models. 

\subsubsection{\label{Subsec:PropertiesA-E}Geometric and Total-Energy Properties of Film Models A-E}

We next investigate the geometries that the molecules assume in the predicted monolayer films A-E after relaxation. A full list of all geometry parameters discussed below is provided for each film in Table~S6 and Table~S7 in the supplementary material, where we define the adsorption distance $d_z$ of a molecule to the substrate as the distance between the $z$-component of the center of mass of the molecule atoms and the plane to the average $z$-position of the top layer of Si atoms (see Figure~S4). %The molecules adsorb at an average distance  of $d_z =10.1$~{\AA} (Pc) and $d_z = 8.9$~{\AA} (Tc) between the average $z$ of the molecular COM and the plane through the first layer of Si~atoms in the surface (see Figure S1 in the SI). 
The Pc molecules in models A-E assume tilt angles $\theta$ of the molecules' long axes to the substrate normal within 2$^\circ$. The same is true for Tc films, except for model D where the angles are on average $\theta = 8^\circ$. Interestingly, these tilt angles value are in line with the GIXRD result for Pc monolayer films on amorphous silicon oxide\cite{mannsfeldBaoAdvMater2009}, two layered Pc films on self-assembled membranes\cite{yangBaoJACS2005} and with the tilt angle suggested for the thin-film phase of Pc\cite{yoshidaSato2007,schiefer2007, ambroschdraxlMeisenbichlerNewJPhys2009}. 
The average edge-to-face or herringbone angle $\omega$ for models B, C, D and E lies between $51^\circ$ and $54^\circ$, again in close agreement to the experimental value for Pc on amorphous silicon oxide\cite{mannsfeldBaoAdvMater2009}. In contrast, the much more compressed small-cell model A shows a slightly smaller average herringbone angle of $\omega = 46^\circ$ for both Pc and Tc.

While amorphous silicon oxide is of course a different substrate than H/Si(111), the agreement observed here regarding tilt angles is consistent with the idea that the films interact only weakly with these substrates and the detailed film structure is largely determined by intermolecular interactions within the films. The small value of $\theta$ indicates that the higher-density monolayer film model reflects the experimentally suggested geometry better than Model $\Phi$, which was built based on the experimental supercell lattice parameters for Pc\cite{nishikataPRB2007}.  While speculative, it seems possible that the PBE+TS density functional used to construct the free-standing film model is slightly too attractive, necessitating a smaller lattice parameter than the experimental film. In this scenario, the correct molecular tilt would be obtained for the denser model. In contrast, when using the experimental lattice parameter (Model $\Phi$), the molecules in the film would tilt too much, in order to improve their overall packing as dictated by DFT-PBE+TS, as we observe in Model $\Phi$.

A comparison of the energy per molecule $\Delta E$ for the combined film-on-substrates models after geometry optimization is given in Figure~\ref{Fig:PcTcHSi111:coincidencepattern}e. The adsorption energies for the low-strain unit cells B-E are within 0.05~eV per molecule of each other. Film A, where the film is more compressed, is less favorable by about 0.15~eV per molecule. Model $\Phi$ that is based on experimental lattice parameters for Pc\cite{nishikataPRB2007}, where the film is stretched, shows the same (even very slightly more favorable) overall adsorption energy compared to models B-E. This indicates that the unknown, hypothetical exact film-substrate interface with minimum total energy at the Born-Oppenheimer surface (no finite- or zero-$T$ vibrational effects) might be somewhere in between, i.e., slightly stretched compared to the free-standing monolayer film model. 

\subsubsection*{Electronic Structure of Models A--E}

\begin{figure}[h]
\centering
\includegraphics[width=0.8\textwidth]{{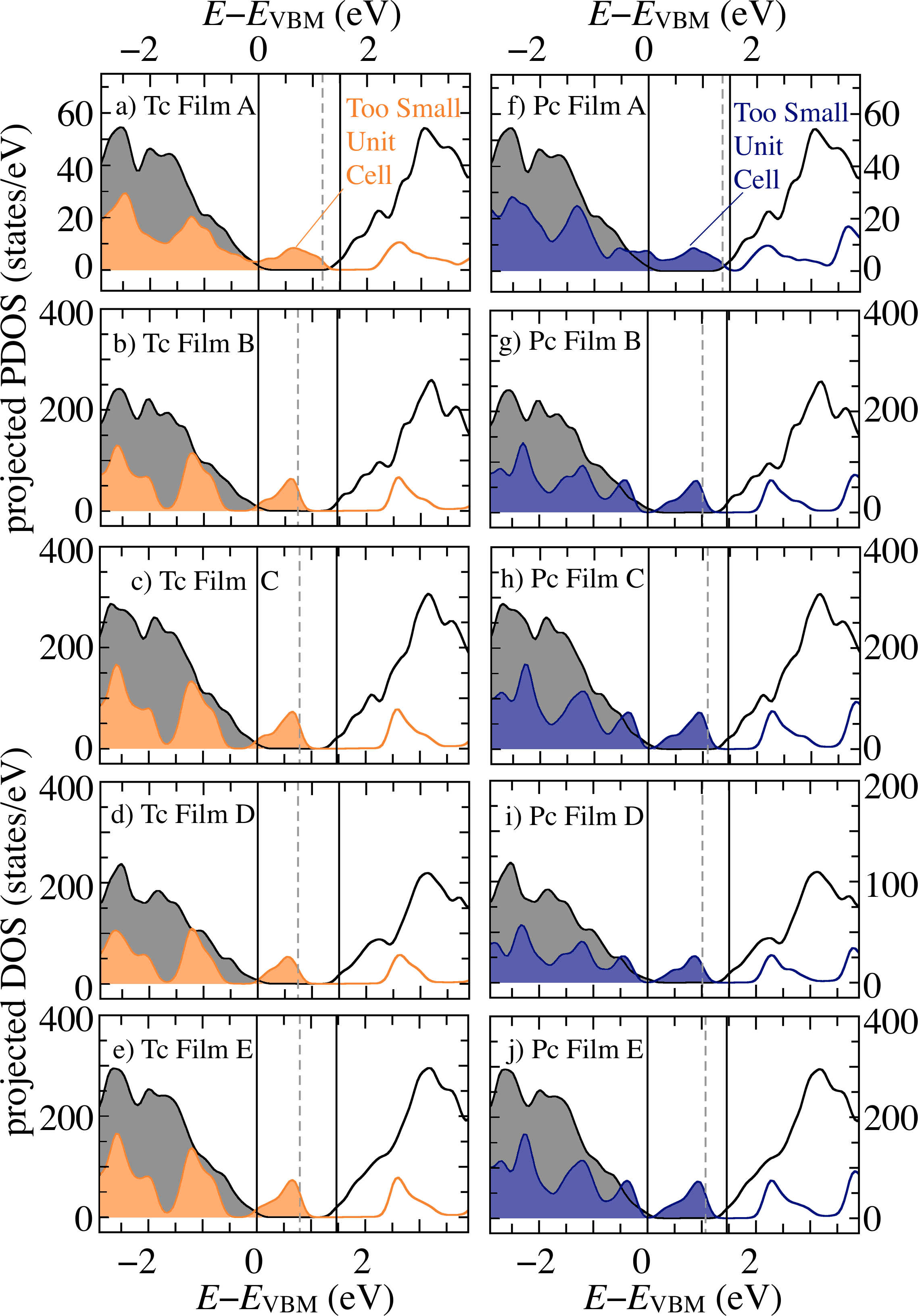}}
\caption{\label{Fig:PcTc:film:PDOS:all}The DFT-HSE06 DOS projected onto the  substrate atoms (black) the Tc (orange) and Pc (blue) film atoms for a) Tc film A b) Tc film B c) Tc film C, d) Tc film D, e) Tc film E, f) Pc film A, g) Pc film B, h) Pc film C, i) Pc film D,  j) Pc film E. Filled areas indicate occupied levels, empty areas unoccupied levels. The black solid lines mark the position of the substrate's HOMO (set as the energy zero in each case) and LUMO. The dashed grey lines mark the position of the electron chemical potential. Due to the artificial Gaussian broadening used for the PDOS (0.1~eV), the peaks of the overall HOMO are visually broadened above the electronic chemical potential level. Actual fractional occupations in the self-consistent DFT calculations are only found in films A, not in B-E.} 
\end{figure}

Figure~\ref{Fig:PcTc:film:PDOS:all} shows the densities of states projected onto the substrate and organic monolayer for Tc (a--e) and Pc (f--j) for the combined film-on-substrate models A--E. The same general observations hold for supercell models B--E. Similar to Model $\Phi$ (Fig.~\ref{Fig:Pc:Phi:PDOS}), the LUMOs of both acene monolayer films are within the substrate's conduction band. Likewise, the films' HOMO levels (dashed grey lines in Fig.~\ref{Fig:PcTc:film:PDOS:all}) are in the substrate's band gap. The HOMO-LUMO gap of the isolated standing monolayers is on average  $1.75$~eV for Tc models B--E and $1.12$~eV for Pc models B--E, slightly higher than the 1.05~eV found for Model $\Phi$. The level alignment is qualitatively the same type II heterojunction as found for $\Phi$. Importantly, as noted above for model $\Phi$, the remaining uncertainty of the substrate band gap of $\approx$0.3~eV due to the finite thickness of the Si slab is not large enough to alter this qualitative conclusion for models B--E. While the densities of states shown are artificially broadened using a Gaussian broadening of 0.1~eV to obtain smooth curves, the actual predicted differences between the film HOMO (dashed grey lines) and the substrate CBM (upper black lines) are visibly larger in these cases in Fig.~\ref{Fig:PcTc:film:PDOS:all}.
 
In contrast, for the much more compressed film of the small-cell approximant unit cell A, more significant changes of the electronic structure are observed. First, its molecular HOMO seems broadened. As a consequence, the HOMO-LUMO bandgap for both Tc ($1.10$~eV for model A) and Pc ($0.44$~eV for model A) has closed considerably compared to the less strained models. More significantly, the overall band gap of the combined film-substrate system -- which is still present for Tc for Pc in models B-E -- is more than half for model A for both Tc (0.33~eV) and for Pc (0.14~eV). For Pc, this means that the band gap closes almost entirely (see also Table~S8 in the SI). In view of these changes, the large-strain approximant cell A is not a safe model to predict the electronic structure of this organic-inorganic hybrid system. 

\begin{figure}[h]
\centering
\includegraphics[width=0.99\textwidth]{{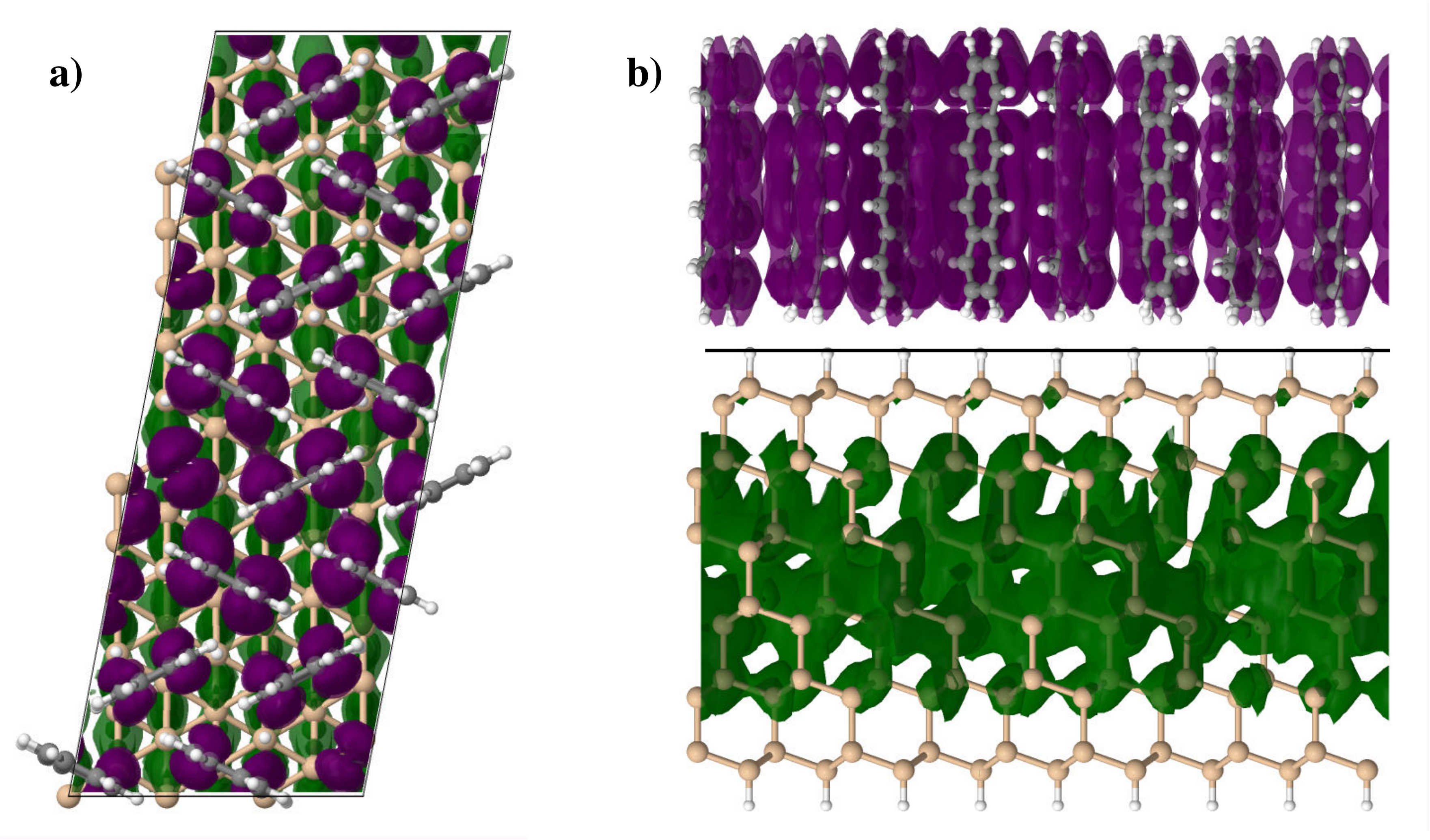}}
\caption{\label{Fig:PcTc:Eldensity}Visualization of the superposition of the eigenstates that contribute to the frontier orbitals of the HOMO (purple, three eigenstates) and LUMO (green, two eigenstates) at the $\Gamma$-point in tetracene film B at HSE06 level of theory, viewed along the $z$-axis (a) and the $x$-axis (b) of the computational supercell.  The HOMO and the LUMO orbitals were visualized with Jmol\cite{jmol} using a cutoff of 0.0004 e/\AA$^3$.}  
\end{figure}

In Figure~\ref{Fig:PcTc:Eldensity}, we visualize the frontier orbitals, i.e., the HOMO (purple) and the LUMO (green), at the $\Gamma$-point for the example of Tc film B in Figure \ref{Fig:PcTc:film:PDOS:all}c for a superposition of three and two degenerate eigenstates (within 10\,meV of one another) for the HOMO (Figure~\ref{Fig:PcTc:Eldensity}a) and LUMO (Figure~\ref{Fig:PcTc:Eldensity}b), respectively. The localization of the eigenstates does not change for the following five eigenstates. In agreement with the PDOS analysis above, we find that the nearly degenerate eigenstates that form the HOMO are delocalized on the molecular film. The eigenstates contributing to the LUMO are similarly delocalized on the more bulk-like atoms in the middle of the Si-slab, in agreement with the observation that the onset of the band gap is formed by contributions from bulk layers of the slab (see Figure~S5 in the SI). The HOMO and LUMO are hence indeed spatially removed from one another, with what appears to amount to a small barrier formed by the interface itself. This supports the idea that a charge separation of electrons and holes could occur in a type II heterojunction-like manner at the interface between Tc and Pc monolayer films and H/Si(111). 

In section~\ref{Subsec:Phi}, we concluded that a transfer of electrons generated by split triplet excitons to the substrate CBM seemed possible in view of our computed DOS for Model $\Phi$. This conclusion was based on the expected CBM level for a slab of converged thickness and literature values of singlet and triplet energies in Pc. Within the uncertainties of both the HSE06 density functional and our model (restricted slab thickness), the same observations remain true for the Pc models B--E. For Tc, the experimentally measured triplet energy is $\sim1.25$~eV\cite{tomkiewiczJCP1971, vauvelBaesslerMolCrystLiqCryst1971, geacintov1969} and that of the singlet 2.32~eV\cite{tomkiewiczJCP1971}. With these values, the approximate triplet level in the Tc films in our models is again slightly above the inorganic LUMO (CBM). Thus, singlet and triplet excitons in both Pc and Tc could potentially dissociate at the heterojunction and pass electrons into the inorganic substrate, while holes would remain on the film.

Finally, based on UV photoelectron spectroscopy measurements of $n$-doped, bare H/Si(111) samples and on H/Si(111) samples coated with 10~nm or more Tc, MacQueen \textit{et al.}\cite{macqueenLipsMaterHoriz2018} suggest a type I level alignment with the HOMO of the Tc film, lying $\approx150$~meV below the VBM of the Si substrate. These observations are at variance with our computational observations for much thinner films, which account for the levels in a single system, i.e., including the electrostatics at the interface.\cite{Nabok2019} It seems possible that differences in film thickness and substrate doping might lead to different level alignment between our computational results and the experimental observation by MacQueen \textit{et al.} On the other hand, our calculations are performed for the combined film-substrate system, not for separate film and substrate systems that were used to deduce alignments in the experimental study. Encouragingly, for the combined film-substrate system, MacQueen \textit{et al.}\cite{macqueenLipsMaterHoriz2018} observe a hole transfer from the substrate to the Tc film that is consistent with the conclusions drawn from our model system. Finally, their suggestions of a triplet exciton level roughly degenerate with the LUMO of  the H/Si(111) is also consistent with our qualitative picture.

\clearpage

\section{Conclusion}
In conclusion, we present a complete computational study, using hybrid density-functional theory, of the electronic level alignment of Pc and Tc molecules and monolayer films at the intrinsic, i.e., undoped, H/Si(111) interface. As part of our study, we describe the impact of two different approaches to construct appropriate, commensurate supercell models for weakly bonded crystalline thin films on a crystalline substrate with a different lattice parameter:
\begin{itemize}
\item The first approach, shown for Pc, relies on available experimental input data as far as possible, particularly regarding the overall film lattice parameter and the molecular orientation (herringbone) inside the film. The remaining (unknown) geometric parameters of the film are obtained by structure optimization using standard van der Waals corrected semilocal density-functional theory. The resulting model (``Model $\Phi$'' in the text above) achieves a plausible description of the electronic levels at the surface, but yields intra-film geometry parameters that differ from experimentally known molecular orientations and tilt angles in Pc monolayer films on other substrates\cite{mannsfeldBaoAdvMater2009}. Conceivably, the experimental lattice parameter is larger than the optimum lattice parameter that would be achieved by the density functional, causing other structural distortions (molecular tilts) in the theoretical result to compensate for the small systematic errors of the overall film lattice parameter in the theory.
\item The second approach relies on using fully computationally obtained film and substrate lattice parameters, inferred from a free-standing model for the organic film, which are then joined to form commensurate supercells of varying size and strain. This approach yields molecular orientations in Pc films that are quite similar to available experimental data. A consistent description of electronic levels is achieved based on all commensurate film models considered that are large enough (up to 1,192 atoms) to exhibit low internal strain. In contrast, the simplest and smallest commensurate supercell model, with $\approx$10\% compressive strain, leads to noticeable distortions of the electronic structure.
\end{itemize}
Our investigation further reveals type-I like level alignments for the isolated Pc and Tc molecules, which prefer lying orientations on the substrate. Films of Pc and Tc exhibit standing molecular geometries and type-II like level alignments with the substrate. According to these results, it should be possible to separate carriers at the organic-inorganic interface, with electrons passing into the substrate and holes remaining on the films. In particular, based on our results and on known exciton energies in the literature\cite{geacintovStrom1971, burgos1977, vilarSchottChemPhysLett1983, deCheveigneDefourneauPRB1977, tomkiewiczJCP1971, vauvelBaesslerMolCrystLiqCryst1971, geacintov1969}, triplet excitons should also be able to split at the interface and inject electrons into the substrate. We finally note that our computationally predicted type-II level alignment in Tc films on intrinsic H/Si(111) substrates is not in agreement with a recent UV photoelectron study of thicker Tc films on n-doped H/Si(111).\cite{macqueenLipsMaterHoriz2018} The latter study concludes type-I level alignment between the transport levels of the Tc film and the Si substrate, albeit from separate photoemission spectroscopy measurements of the clean substrate and the film. Nevertheless, in the same study, hole extraction from the substrate to the organic films is still possible in the combined film-substrate system, i.e., in principle in line with our result of type-II level alignment. While the origins of the observed differences are not clear, changes due to substrate doping are one possibility. It would be interesting to extend our methodology to consider semiconductor substrates with controlled doping densities, i.e., Fermi level, in future work.

\ack
S.M.J. thanks the Deutsche Forschungsgemeinschaft (DFG, German Research Foundation) for a postdoctoral fellowship, grant number 393196393. 
In addition, this work was funded by the Deutsche Forschungsgemeinschaft (DFG) - Projektnummer 182087777 - SFB 951. V.B. acknowledges support from N.S.F. Award DMR-1729297.

\bibliographystyle{iopart-num}
\section*{References}
\bibliography{paper_HIOS}
%\endbib
\end{document}

\begin{titlepage}
    \begin{center}
        \vspace*{1cm}
 
        \huge
        \textbf{Supplementary Information}

        \vspace*{0.5cm}

        \textbf{Pentacene and Tetracene Molecules and Films on H/Si(111): Level Alignment from Hybrid Density Functional Theory}
 
        \vspace{1.5cm}
 		%\normalsize
 		\Large
        Svenja M. Janke$^{1,2}$, Mariana Rossi$^{2,3}$, Sergey Levchenko$^{4,2}$, Sebastian Kokott$^{2}$, Matthias Scheffler$^2$ and Volker Blum$^{1,5}$\\
        \vspace{1cm}
        $^{1}$ \textit{Department of Mechanical Engineering and Materials Science, Duke University, Durham, NC, USA}\\
        $^{2}$ \textit{Fritz Haber Institute of the Max Planck Society,  Berlin, Germany}\\
        $^{3}$ \textit{Max Planck Institute for Structure and Dynamics of Matter, Hamburg, Germany} \\
        $^{4}$ \textit{Skolkovo Institute of Science and Technology, Moscow, Russia} \\
        $^{5}$ \textit{Department of Chemistry, Duke University, Durham, NC, USA}
        %\textbf{Author Name}
 
        \vfill
 
        \vspace{0.8cm}
 
    \end{center}
\end{titlepage}

\section*{Bulk Pentacene}
\begin{table}[h]
\centering
\caption*{\label{Tab:cell:Pc:PBETSPBEMBD:bulk}\textbf{Table S1}: Comparison of the lattice parameters and angles of the crystal polymorph of pentacene  obtained at 90\,K and 300\,K from x-ray diffraction~\cite{mattheusThomas2001} to values obtained after relaxation with DFT-PBE and Tkatchenko-Scheffler (TS)~\cite{tkatchenko2009} as well as with many-body dispersion (MBD)~\cite{ambrosetti2014} van der Waals (vdW) correction with ``intermediate'' settings. All residual forces on the atoms and lattice vectors in the computational structures are below 0.005\,eV/{\AA}. The computational energy minimization of the structure used the lattice parameters and atomic positions as supplied in the $T$=90\,K cif-file of the experimental publication~\cite{mattheusThomas2001} as its initial guess. All atomic positions and lattice parameters were allowed to relax. The values obtained with DFT-PBE+TS are in good agreement with the DFT-PBE+MBD and experimental values.}
\begin{tabular}{lccccccc} 
 \hline
&$\mathbf{a}_1$ (\AA)	&$\mathbf{a}_2$ (\AA)	&$\mathbf{c}$ (\AA)&$\alpha$ ($^\circ$)& $\beta$ ($^\circ$)& $\gamma$ ($^\circ$)\\
 \hline\hline
 DFT-PBE+TS	&6.11	&7.63	&14.59	& 78 	& 87	& 85 \\
 DFT-PBE+MBD&6.25	& 7.67	& 14.38 	& 77 	& 89 	& 84	\\
 \hline
 90\,K~\cite{mattheusThomas2001}&6.239(1)& 7.636(1) &14.330 (2)& 76.978(3) & 88.136(3) &84.415(3)\\
 300\,K~\cite{mattheusThomas2001}&6.266(1) & 7.775(1) &14.530(1) & 76.475(4) &87.682(4) &84.684 (4)\\
 \hline
\end{tabular}
\end{table}
\clearpage

\section*{Bulk Tetracene}
\begin{table}[h]
\centering
\caption*{\label{Tab:cell:Tc:PBETSPBEMBD:bulk}\textbf{Table S2}: Comparison of the lattice parameters and angles of a tetracene crystal measured by x-ray diffraction~\cite{holmesVollhardt1999,sondermannBasslerJPC1985} to values obtained after relaxation with DFT-PBE+TS and DFT-PBE+MBD at ``intermediate'' settings. All residual forces on atoms and lattice vectors were below 0.005\,eV/{\AA}. The relaxation used the lattice parameters and atomic positions as supplied in the cif-file of the experimental publication~\cite{holmesVollhardt1999}, CSD 114446~\cite{GroomWard_ActaCrystB2016} as initial guess. All atomic positions and lattice parameters were allowed to relax.}
\begin{tabular}{lccccccc} 
 \hline
&$\mathbf{a}_1$ (\AA)	&$\mathbf{a}_2$ (\AA)	&$\mathbf{c}$ (\AA)&$\alpha$ ($^\circ$)& $\beta$ ($^\circ$)& $\gamma$ ($^\circ$)\\
 \hline\hline
DFT-PBE+TS										&5.97 	&7.76 	&12.97	& 77	& 73	& 86\\
DFT-PBE+MBD										&5.98 	&7.82 	&12.95 	& 77 	& 72 	& 86\\
\hline
175\,K~\cite{holmesVollhardt1999}	&6.0565(9)		&7.8376(11)  &13.0104(18) 	& 77.127(2) 	& 72.118(2) 	& 85.792(2)\\ % temperature checked by VB.
140~\cite{sondermannBasslerJPC1985} &5.99(2)&7.77(4) &13.2(10) &102.4(5) &113.8(9) &86.0(4)\\
293\,K~\cite{sondermannBasslerJPC1985} &6.065(2) & 7.915(5) &13.445(12) &101.10(6) & 113.31(9) &85.91(4)\\
 \hline
\end{tabular}
\end{table}
\clearpage

\section*{Bulk Si}
\begin{table}[h]
\centering
\caption*{\label{Tab:latticeconstant}\textbf{Table S3:} Predicted lattice parameters for bulk Si using different density functionals and van der Waals corrections with a ($12\times 12\times 12$) k-point mesh and ``tight settings''. Minimum-energy lattice parameters at the Born-Oppenheimer surface were extracted by fitting total energies for different volumes to the Birch-Murnaghan formula~\cite{murnaghanPNAS1944, birchPR1947}.}
\begin{tabular}{lc} 
 \hline
  Functional& $a_0$ (\AA)  \\  
 \hline\hline
DFT-PBE&5.472\\
%PBEsol&5.436\\
DFT-PBE+TS&5.450\\
DFT-PBE+MBD&5.436\\
%PBE0+MBD&5.407\\
HSE06+MBD&5.414\\
HSE06+TS&5.426\\
Lit. (RT)~\cite{crc2005, kittel2005}&5.431\\
Lit. (6.4\,K)~\cite{batchelderSimmonsJCP1964}&5.419\\
 \hline
\end{tabular}
\end{table}
\clearpage

\section*{H/Si(111) Band Gap Convergence with Slab Thickness}
\begin{figure}[h]
\centering
\includegraphics[width=0.99\textwidth]{{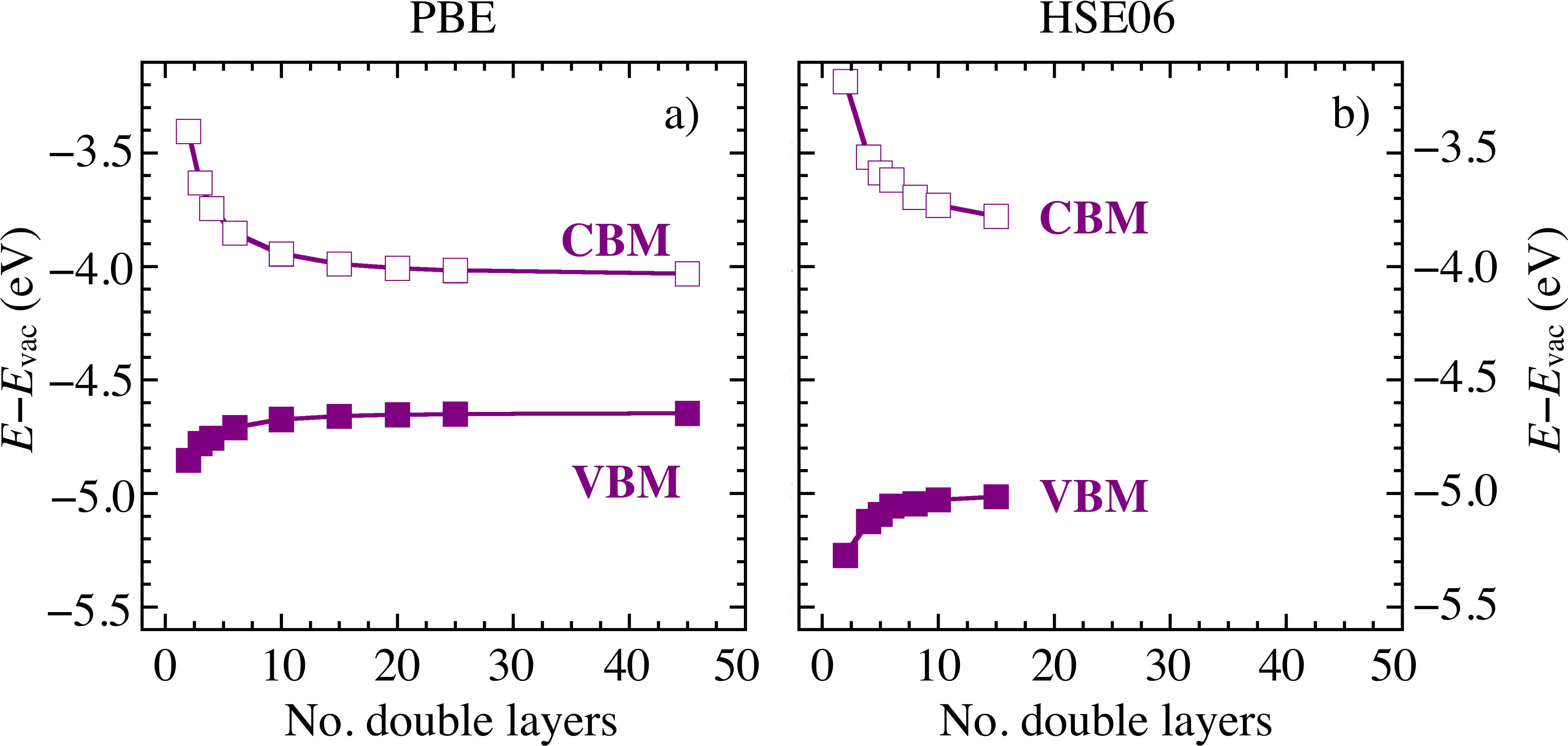}}
\caption*{\label{Fig:slabconergence}\textbf{Figure S1:} Variation of the valence band maximum (VBM, filled purple squares) and the conduction band minimum (CBM, open purple squares) of the H/Si(111) slab models as a function of the number of double layers. The energy axis values are referenced to the vacuum level ($E-E_\mathrm{vac}$). a) DFT-PBE results. b) DFT-HSE06 results from surfaces relaxed with DFT-PBE+TS and HSE06+TS. We relaxed  ($1\times 1\times n$) H/Si(111) slabs as a function of the number of Si double layers $n$ in $z$-direction, using a ($12\times 12\times 1$) k-point mesh and a vacuum layer thickness of 200\,\AA, until the minimum force was below 0.01\,eV/\AA~with ``tight'' numerical settings of FHI-aims. The atomic positions of the top and bottom two Si double layers and the H~atoms were relaxed while all others were kept fixed during the relaxation. The slow convergence of the band gap with the thickness of the slab has been observed previously for various Si surfaces~\cite{liGalliPRB2010, sagisakaBowlerJPhyCondensMat2017, delleySteigmeierApplPhysLett1995, scherpelzGalliPRMater2017} and is explained by a quantum well behavior due to confinement of the electronic eigenstates in thin slabs~\cite{scherpelzGalliPRMater2017, delleySteigmeierApplPhysLett1995, fischettiJiseok2011}. The DFT-HSE06 bulk band gap of 1.165\,eV (``tight'' settings) is close to the experimental value of 1.17\,eV at $T$=0\,K~\cite{kittel2005}.}
\end{figure}

\clearpage

\section*{Dilute Limit of Acene Adsorption on H/Si(111)}
\begin{figure}[h]
\centering
	\includegraphics[width=0.99\textwidth]{{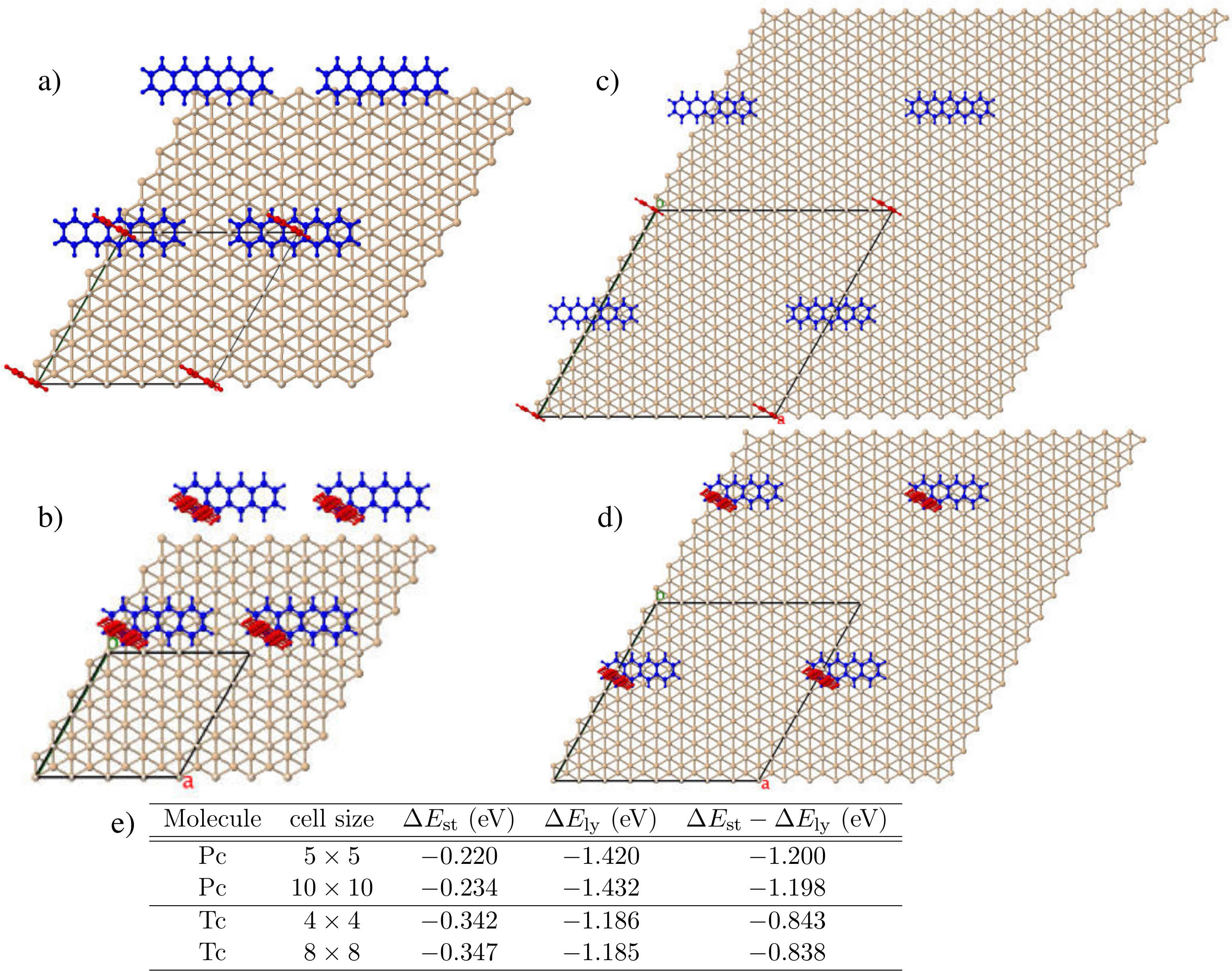}}
\caption*{\label{Fig:Supplementary:supercell}\textbf{Figure S2:} Images of lying (blue) and standing (red) acenes adsorbed on H/Si(111) in the unit cells considered for the dilute limit in this work. We chose a cell size in $xy$-plane of ($5\times 5$) for Pc (a) and ($4\times 4$) for Tc (b) to isolate each molecule from its periodic images. If the cell size is doubled (c, d), the energy per molecule remains within a few meV of that of the smaller cell size (e, Table), both for a lying molecule ($\Delta E_\mathrm{ly}$) and a standing molecule ($\Delta E_\mathrm{st}$). Total  energy calculations were performed using DFT-PBE+TS and ``intermediate'' settings. For the geometries in a--e, a previously relaxed $(1\times 1\times 10)$ double layer unit cell was repeated to form the different underlying slabs. For the positions of the molecules, we selected one molecule in a lying and one in a standing orientation from the structure search in Figure 3 of the main text. We then transferred the molecular positions to the slab models shown in this figure, keeping the distance to the average plane through the surface hydrogen atoms in the original structure.}
\end{figure}
\clearpage

\section*{Computational Parameters for Interface Models of the Dilute and Monolayer Limit}
\begin{table}[h]
\centering
\caption*{\label{Tab:Denselimit:kmesh:I}\textbf{Table S4:} For different models considered in the text, the table shows the unit cell area, total number of atoms (film and substrate) per unit cell, and k-point meshes used for structure relaxation (DFT-PBE+TS) and electronic structure calculation (HSE06). The HSE06 calculations were performed using the structures obtained with DFT-PBE+TS. ``Intermediate'' settings were used.}
\begin{tabular}{lccccc} 
 \hline
Model& area (\AA$^2$)& No. Tc&  No. Pc& DFT-PBE+TS   & HSE06\\
& & +H/Si(111) & +H/Si(111)  & k-point mesh & k-point mesh\\  
&& atoms per cell & atoms per cell\\
 \hline\hline
\multicolumn{3}{l}{Dilute limit} \\
\hline
Tc& 206 & 382&&($2\times 2\times 1$)&($5\times 5\times 1$)\\
Pc& 322 & &586&($2\times 2\times 1$)&($5\times 5\times 1$)\\
 \hline
\multicolumn{3}{l}{Monolayer limit}\\
 \hline
A& 77   &  204&228&($3\times 5\times 1$)&($4\times 7\times 1$)\\
B& 347 & 858&954&($6\times 2\times 1$)&($3\times 1\times 1$)\\
C& 424 & 1062&1182&($6\times 2\times 1$)&($3\times 1\times 1$)\\
D& 309 & 756&840&($6\times 2\times 1$)&($3\times 1\times 1$)\\
E& 424 & 1062&1182&($6\times 3\times 1$)&($4\times 2\times 1$)\\
$\Phi$ &283 & &  740 &($6\times 2\times 1$)&($6\times 2\times 1$)\\
\hline
\end{tabular}
\end{table}
\clearpage

\section*{Free-Standing Monolayer Relaxation Geometries}
\begin{figure}[h]
\centering
\includegraphics[width=0.99\textwidth]{{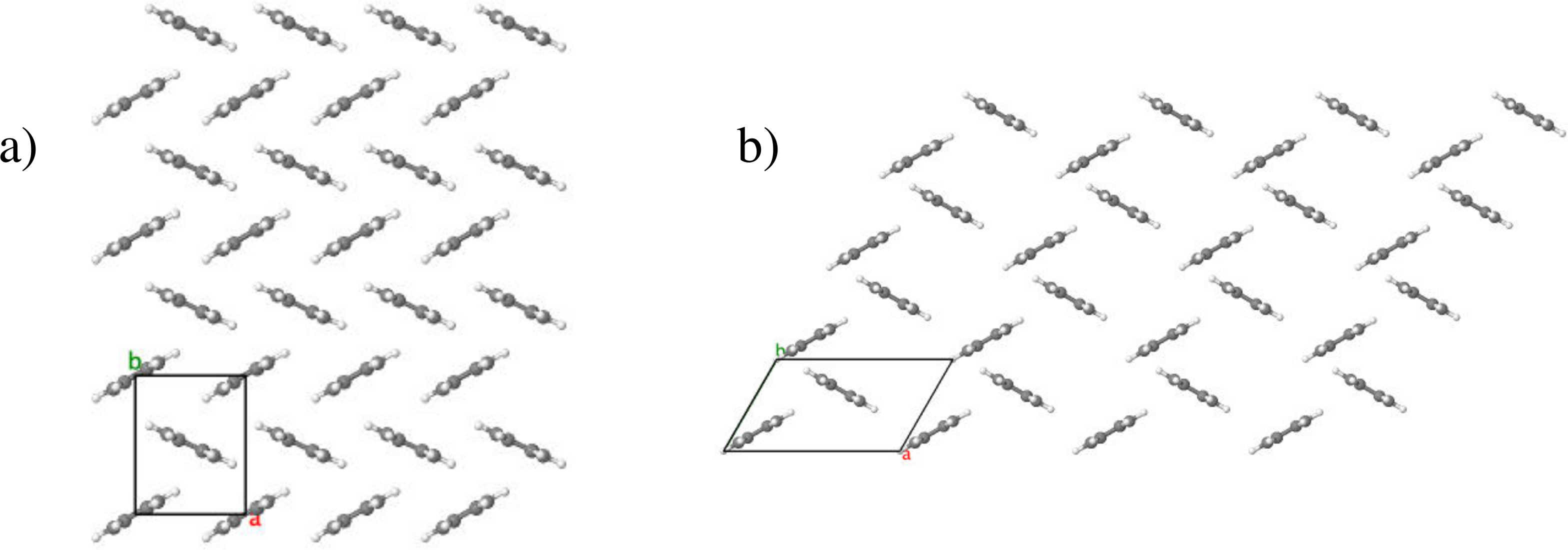}}
\caption*{\label{Fig:TcmonolayerStructure}\textbf{Figure S3:} Initial geometry before relaxation of a Pc (a) and Tc (b) monolayer in vacuum viewed along $\mathbf{c}$-direction. The unit cell used in the relaxation is indicated by a black frame. For Pc, the initial unit cell was based on the unit cell I structure suggested by Nishikata  \textit{et al.}~\cite{nishikataPRB2007}. For Tc, the initial unit cell was based on the observation that Tc molecules assume a standing orientation and a herringbone pattern on weakly interacting surfaces~\cite{shi2006}. Relaxations were performed with PBE+TS and ``intermediate'' settings until all residual forces on atoms and lattice vectors were below 0.005~eV/{\AA}.}
\end{figure}
\clearpage

\section*{Free-Standing Monolayer Lattice Parameters and Energies}
\begin{table}[h]
\centering
\caption*{\label{Tab:filminvacuum:unit cell}\textbf{Table S5:} In-plane lattice parameters of the Pc and Tc unit cells after relaxation, calculated using ``intermediate'' settings with DFT-PBE  and TS~\cite{tkatchenko2009} as well as with MBD~\cite{ambrosetti2014} van der Waals (vdW) corrections. All cells were rotated to have one of their lattice parameters parallel to the $x$-axis. Within the cells, all molecules assumed a herringbone pattern after relaxation. For Pc, unit cells were based on unit cell I  determined for Pc dendrites on H/Si(111)~\cite{nishikataPRB2007}.
%The structures for Tc (90$^\circ$) and (60$^\circ$) are identical. 
For Tc, a unit cell was modeled with a herringbone pattern between two standing molecules based on the experimentally described geometry\cite{shi2006, tersigni2006}. For reference, relaxations for Tc were also performed based on the unit cell I used for Pc as a starting point. The resulting lattice vectors are almost identical.
The energy is given with reference to the isolated molecule in vacuum as energy per molecule $\Delta E_\mathrm{mol}$. The two bottom lines show the lattice parameters from the experimental study of Pc dendrites on H/Si(111) at room temperature~\cite{nishikataPRB2007}. Their energy marked ($^\mathrm{a}$) corresponds to the energy after relaxation of the molecular positions with DFT-PBE+TS while the lattice parameters remained fixed to the experimental lattice parameters. The experimental error bars for the unit cell parameters are therefore included in the table and the notation ``exp/TS'' indicates the fact that experimental geometries and computed energies are combined in the lowest two lines.}
\begin{tabular}{lllllll} 
 \hline
&vdW& $c_{11}$ (\AA)& $c_{12}$ (\AA)&$c_{21}$ (\AA)&$c_{22}$ (\AA)& $\Delta E_\mathrm{mol}$ (eV)\\
 \hline\hline
 Pc (unit cell I based)&TS&5.85&0.00&0.00&7.36&$\mathbf{-1.974}$\\
 Tc (unit cell I based)&TS&5.83&0.00&0.02&7.41&$\mathbf{-1.556}$\\
 %Tc (60$^\circ$)&TS&5.921&0.000&0.001&7.333&$\mathbf{-1.554}$\\
 Tc (model)&TS&5.85&0.00&0.00&7.39&$\mathbf{-1.556}$\\
 %Tc (90$^\circ$)&TS&5.917&0.000&$-5.921$&7.340&$\mathbf{-1.554}$\\
 \hline
  Pc (unit cell I based)&MBD&5.89&0.00&0.00&7.40&$\mathbf{-1.588}$\\
 Tc (unit cell I based)&MBD&5.89&0.00&0.00&7.44&$\mathbf{-1.255}$\\ 
 \hline
 \hline
  Pc supercell I&exp/TS&6.02$\pm0.01$&0.00&0.00&7.62$\pm0.01$&$-1.909^\mathrm{a}$\\
  Pc supercell II&exp/TS&5.98$\pm0.01$&0.00&0.00&7.56$\pm0.01$&$-1.932^\mathrm{a}$\\
 \hline
\end{tabular}
\end{table}
\clearpage

\section*{Models of Monolayer Pentacene Interfaces with H/Si(111): Geometries and Energies}
\begin{table}[h]
\centering
\caption*{\label{Tab:Pc:lattice parameters:energy}\textbf{Table S6:} For pentacene, the lattice parameters and angle $\gamma$ between the lattice vectors, the energy per molecule $\Delta E$, the average angle $\theta$ of the molecules to the surface normal, the average herringbone angle $\omega$ and the average distance $d_z$ of the $z$-component of the molecules' center of mass to the plane through the average height of the first layer of Si atoms (see Figure~S4) are shown.}
\begin{tabular}{lrrrrrrrc} 
 \hline
Film&$\mathbf{a}$ (\AA)&	$\mathbf{b}$ (\AA)&$\gamma (^\circ)$&	$\Delta E$ (eV)&$\theta~(^\circ)$&$\omega~(^\circ)$& $d_z$ (\AA)\\
\hline\hline
free-standing&5.85&7.36&90		&\\
\hline
A&5.78	&6.68	&90				&$-2.022$	&$1$	&$46$	&$10.0$\\
B&5.78	&30.59	&79				&$-2.213$	&$1$	&$54$	&$10.2$\\
C&5.78	&37.16	&81				&$-2.221$	&$1$	&$52$	&$10.1$\\
D&13.90&30.10&48				&$-2.175$	&$1$	&$51$	&$10.2$\\
E&5.78	&37.16	&81				&$-2.221$	&$1$	&$52$	&$10.1$\\
$\Phi$&6.13&7.71&87			&$-2.222$	&$22$&$49$	&$9.5$\\
\hline
\end{tabular}
%\end{sidewaystable}
\end{table}
\clearpage
\section*{Models of Monolayer Tetracene Interfaces with H/Si(111): Geometries and Energies}
\begin{table}[h]
\centering
\caption*{\label{Tab:Tc:lattice parameters:energy}\textbf{Table S7:} For tetracene, the lattice parameters and angle $\gamma$ between the lattice vectors, the energy per molecule $\Delta E$, the average angle $\theta$ of the molecules to the surface normal, the average herringbone angle $\omega$ and the average distance $d_z$ of the $z$-component of the molecules' center of mass to the plane through the average height of the first layer of Si atoms (see Figure~S4) are shown.}
\begin{tabular}{lrrrrrrrc} 
 \hline
Film&$\mathbf{a}$ (\AA)&	$\mathbf{b}$ (\AA)&$\gamma (^\circ)$&	$\Delta E$ (eV)&$\theta~(^\circ)$&$\omega~(^\circ)$& $d_z$ (\AA)\\
\hline\hline
free-standing&5.85&7.39&90		&\\
\hline
A&5.78	&6.68	&90				&$-1.625$	&$1$	&$46$	&$8.8$\\
B&5.78	&30.59	&79				&$-1.798$	&$1$	&$54$	&$8.9$\\
C&5.78	&37.16	&81				&$-1.762$	&$1$	&$53$	&$8.9$\\
D&13.90&30.10&48				&$-1.780$	&$8$&$51$	&$8.9$\\
E&5.78	&37.16	&81				&$-1.802$	&$1$	&$53$	&$8.9$\\
\hline
\end{tabular}
%\end{sidewaystable}
\end{table}
\clearpage

\section*{Definition of Monolayer Film Adsorption Distance}
\begin{figure}[h]
\centering
\includegraphics[width=0.60\textwidth]{{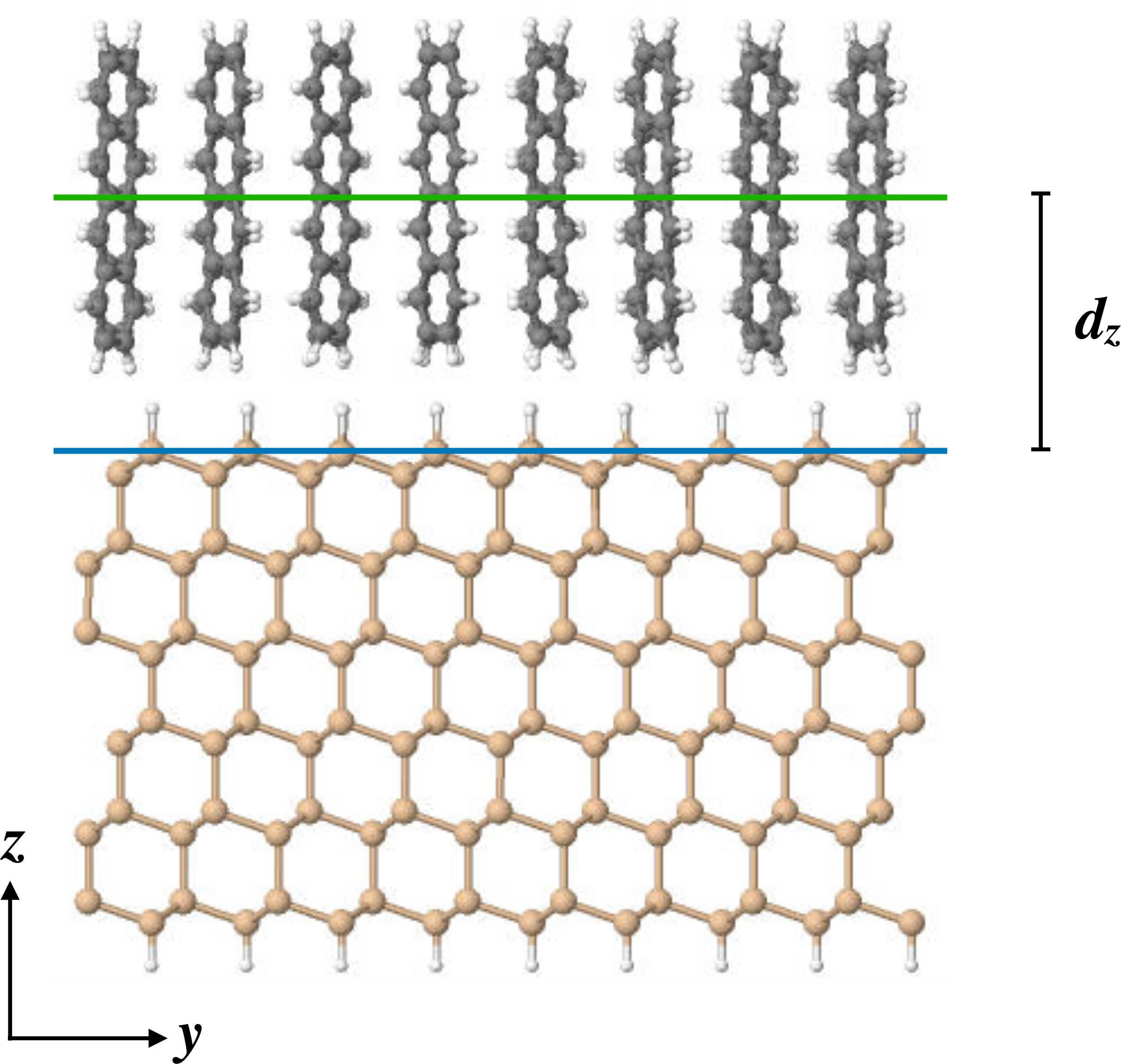}}
\caption*{\label{Fig:slabconergence}\textbf{Figure S4:} Side view of the slab model of a tetracene film on H/Si(111), model B. The adsorption distance $d_z$ of a molecule to the substrate is defined as the distance between the $z$-component of the center-of-mass  of the molecular atoms (green line) and the plane defined by the average $z$-position of the top layer of Si~atoms (blue line).}
\end{figure}
\clearpage

\section*{Frontier Orbital Energies in the Dilute and Monolayer Limits}
\begin{table}[h]
\centering
\caption*{\label{Tab:20181011:Gapsoffilm}\textbf{Table S8:} The DFT-HSE06 energy and the gaps of the VBM and CBM of the substrate and HOMO and LUMO of the film extracted from HSE06 Mulliken analysis, counting all contributions above 0.0025 total occupation. The band gap of the isolated substrate (PBE+TS geometry, calculated with HSE06, ``intermediate'' settings) is 1.465\,eV, the gaps given for ``Tetracene'' and ``Pentacene'' correspond to those  of the isolated molecule  in vacuum.}
\begin{tabular}{c|l|ccc|ccc|c} 
 \hline
Limit &Model&\multicolumn{3}{c}{substrate}&\multicolumn{3}{c}{film}&overall\\
&& VBM& CBM & gap &HOMO &LUMO & gap& band gap\\
&&(eV) & (eV) & (eV)  & (eV)    &(eV)    & (eV)& (eV)\\
\hline
\hline
Isolated&Tetracene&&&&&&$2.269$\\
\hline
\multirow{5}{*}{\shortstack[l]{Monolayer\\Limit}} &A& 
-5.59& -4.08& 1.51& -4.41& -3.31& 1.10& 0.33\\
&B& -5.53&	-4.07&	1.46& -4.86& -3.03& 1.83& 0.79\\
&C& -5.54& -4.07& 1.47& -4.76& -3.07& 1.68& 0.69 \\
%&C& -5.51& -4.02& 1.49& -4.81& -3.07& 1.74& 0.79\\
&D& -5.52& -4.02& 1.50& -4.78& -3.07& 1.71& 0.76\\
&E& -5.54& -4.07& 1.47& -4.80& -3.06& 1.74 & 0.73\\
\hline 
\multirow{2}{*}{\shortstack[l]{Dilute\\Limit}} &Tc lying&
-5.40& -4.07& 1.33& -5.68& -3.51& 2.17& 1.61\\
&Tc standing& -5.42& -4.08& 1.33& -5.42& -3.21& 2.21& 1.33\\
\hline
\hline
Isolated &Pentacene&&&&&&1.696\\
\hline
\multirow{6}{*}{\shortstack[l]{Monolayer\\Limit}} &A& 
-5.67& -4.16& -1.50& -4.30& -3.86& -0.44& 0.14\\
&B& -5.59& -4.13& 1.46& -4.58& -3.42& 1.16& 0.45\\
&C&-5.60& -4.13& 1.47& -4.53& -3.44& 1.09& 0.40\\
&D& -5.66& -4.16& 1.50& -4.66& -3.52& 1.14&  0.50\\
&E&-5.61& -4.14& 1.46& -4.53& -3.44& 1.09& 0.39\\
&$\Phi$&  -5.53& -4.06& 1.46& -4.64& -3.59& 1.05& 0.58\\
\hline
\multirow{2}{*}{\shortstack[l]{Dilute\\Limit}} &Pc lying&
-5.40& -4.07& 1.33& -5.38& -3.74& 1.63& 1.31\\
&Pc standing& -5.42& -4.09& 1.33& -5.14& -3.45& 1.69& 1.06\\
\hline
\end{tabular}
\end{table}
\clearpage
\section*{H/Si(111) Surface Density of States Decomposition}
\begin{figure}[h]
\centering
	\includegraphics[width=0.99\textwidth]{{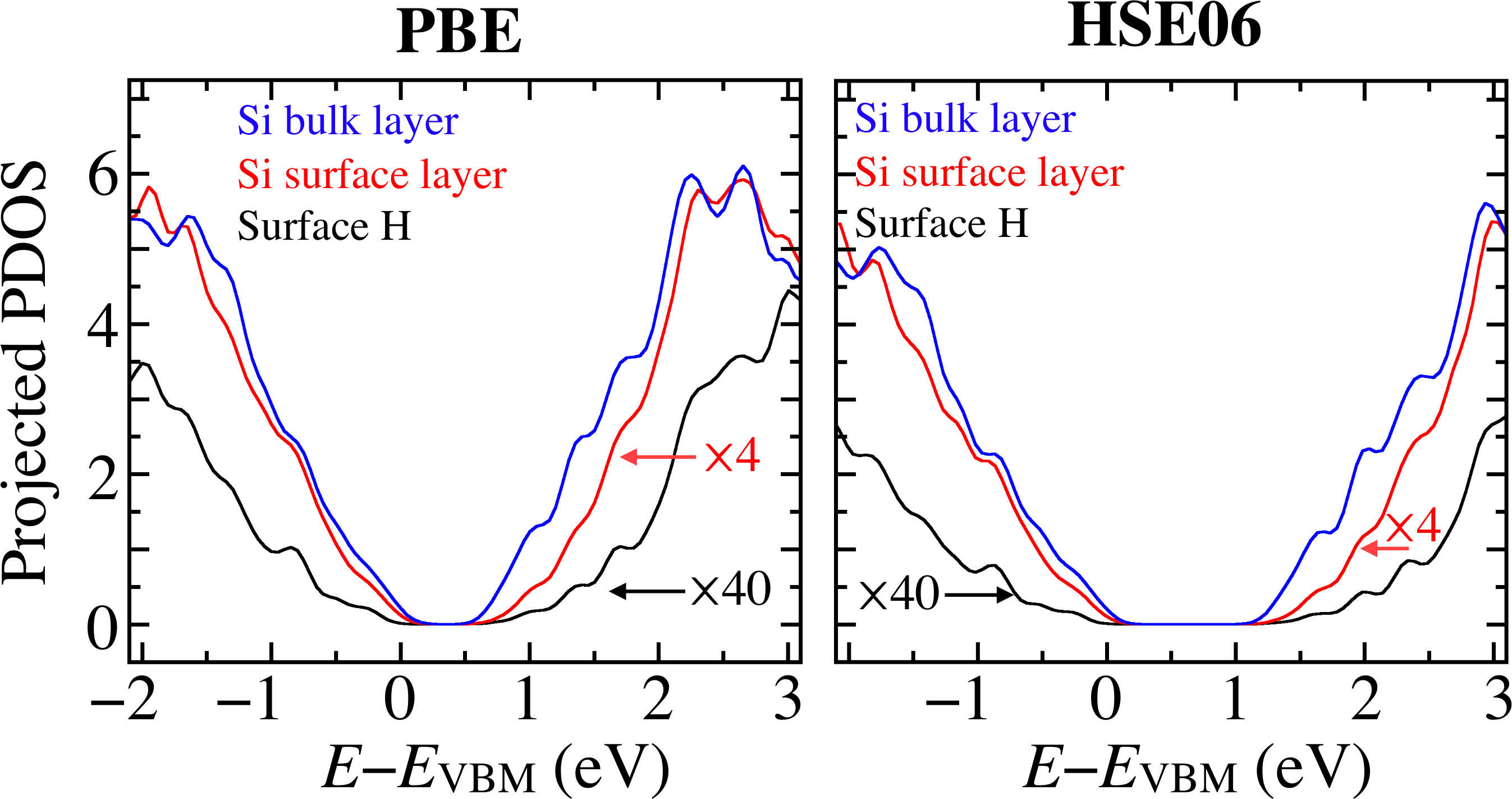}}
\caption*{\label{Fig:Supplementary:bandgap}\textbf{Figure S5:} Projected densities of states onto surface H (black), Si bulk layers (blue) and Si surface layers (red) for 10 Si double layers in H/Si(111) calculated with PBE (left) and HSE06 (right) from surfaces relaxed with DFT-PBE+TS and HSE06+TS at ``intermediate'' settings of FHI-aims. The curves of the surface H and surface Si are scaled by a factor of 40 and 4, respectively. The onset of the band gap is formed by contributions from the bulk layers (blue).}
\end{figure}
\clearpage

\bibliography{paper_HIOS}
%\printindex